\documentclass[10pt,conference,sigconf,review,anonymous,table]{IEEEtran}
\usepackage{amsmath, amssymb}
\usepackage{enumitem}
\usepackage{listings}
\usepackage{booktabs}
\usepackage{comment}
\usepackage{xcolor}
\usepackage{multirow}
\usepackage{subfigure}
\usepackage{caption}
\usepackage{soul}
\usepackage{wrapfig}
\usepackage{seqsplit}
\usepackage{graphicx}
\usepackage{float}
\usepackage{cite}
\usepackage{lipsum}

\usepackage{flushend}

\usepackage[many]{tcolorbox}
\usetikzlibrary{shadows}

\newtcolorbox{shadedbox}{
  drop shadow southeast,
  breakable,
  enhanced jigsaw,
  colback=white,
}

\usepackage[T1]{fontenc}

\definecolor{VeryLightGray}{gray}{0.8}
\definecolor{LightGray}{gray}{0.87}

\newcommand{\tianyin}[1]{{\color{red}{\bf Tianyin:} #1}}
\newcommand{\red}[1]{{\color{red}{\bf #1}}}
\newcommand{\blue}[1]{{\color{black} #1}}
\newcommand{\needCheck}[1]{{\color{red} #1}}
\newcommand{\yuanliang}[1]{{\color{black}{\bf Yuanliang:} #1}}
\newcommand{\haochen}[1]{{\color{black} #1}}

\newcommand{\checked}[1]{{\color{black} #1}}
\newcommand{\Comment}[1]{}
\newcommand{\Space}[1]{}

\newcommand{\mypara}[1]{\noindent \textbf{#1 }}
\newcommand{\myitpara}[1]{\noindent \textit{#1 }}
\newcommand{\CodeIn}[1]{{\small \texttt{#1}}}
\newcommand{\Contrib}[1]{$\star$#1}
\newcommand\blfootnote[1]{%
\begingroup
\renewcommand\thefootnote{}\footnote{#1}%
\addtocounter{footnote}{-1}%
\endgroup
}

\newcommand{\scdi}{configuration design and implementation}

\usepackage[hidelinks]{hyperref}
\newcommand{\myminus}{-}
\newcommand{\HDFS}{HDFS}
\newcommand{\HBase}{HBase}
\newcommand{\HBASE}{HBase}
\newcommand{\SPARK}{Spark}
\newcommand{\Cassandra}{Cassandra}
\newcommand{\CASSANDRA}{Cassandra}

\newcommand{\hdfs}{HDFS}
\newcommand{\hbase}{HBase}
\newcommand{\spark}{Spark}
\newcommand{\cassandra}{Cassandra}
\newcommand{\subjectHeader}{Subject}
\newcommand{\descHeader}{\#Description}
\newcommand{\paramsHeader}{\#Params}
\newcommand{\studiedCommitsHeader}{\#StudiedC}
\newcommand{\allCommitsHeader}{\#AllC}
\newcommand{\confcommits}{configuration-related commits}

\newcommand{\Confcommits}{Configuration-related commits}
\newcommand{\header}[1]{\textbf{\textsc{#1}}}

\DeclareCaptionFont{mysize}{\fontsize{9}{8}\selectfont}
\newcommand{\ttvar}[1]{\texttt{\expandafter\dottvar\detokenize{#1}\relax}}
\newcommand*\dottvar[1]{\ifx\relax#1\else
  \expandafter\ifx\string.#1\string.\allowbreak\else#1\fi
  \expandafter\dottvar\fi}

\definecolor{darkgreen}{RGB}{0,87,0}
\definecolor{amber}{rgb}{1.0, 0.75, 0.0}
\definecolor{ForestGreen}{rgb}{0.13, 0.55, 0.13}

\lstdefinelanguage{diff}[]{Java}{
  morekeywords={assert},
  morekeywords={require},
  morekeywords={checkArgument},
  morecomment=[f][\color{blue}]{@@},     
  morecomment=[f][\color{red}]-,         
  morecomment=[f][\color{magenta}]{---}, 
  morecomment=[f][\color{magenta}]{+++},
  numbers=left,
  basicstyle={\scriptsize\ttfamily},
  numbersep=6pt,
  numberstyle=\tiny\ttfamily\color{gray},
  keywordstyle=\color{blue},
  commentstyle=\color{gray},
  stringstyle=\color{purple},
  breaklines=true,
  escapeinside={(*@}{@*)},
  showstringspaces=false,
  tabsize=2,
  frame=tb,
  xleftmargin=8pt,
}

\renewcommand{\IEEEbibitemsep}{0pt plus 0.5pt}
\makeatletter
\IEEEtriggercmd{\reset@font\normalfont\fontsize{7.9pt}{8.40pt}\selectfont}
\makeatother
\IEEEtriggeratref{1}

\begin{document}
\pagestyle{empty}
\title{An Evolutionary Study of Configuration Design\\ and Implementation in Cloud Systems}

\author{\IEEEauthorblockN{Yuanliang Zhang\IEEEauthorrefmark{1}\IEEEauthorrefmark{2},
Haochen He\IEEEauthorrefmark{1}, Owolabi Legunsen\IEEEauthorrefmark{3},
Shanshan Li\IEEEauthorrefmark{1}, Wei Dong\IEEEauthorrefmark{1}, Tianyin Xu\IEEEauthorrefmark{2}}
\IEEEauthorblockA{
\IEEEauthorrefmark{1}National University of Defense Technology, Changsha, Hunan, China\\
\IEEEauthorrefmark{2}University of Illinois at Urbana-Champaign, Urbana, IL 61801, USA\\
\IEEEauthorrefmark{3}Cornell University, Ithaca, NY 14850, USA\\
\{zhangyuanliang13, hehaochen13, shanshanli, wdong\}@nudt.edu.cn, legunsen@cornell.edu, tyxu@illinois.edu
}
}

\maketitle

\begin{abstract}

Many techniques were proposed for detecting software
  misconfigurations in cloud systems
  and for diagnosing unintended behavior
  caused by such misconfigurations.
Detection and diagnosis are steps in
  the right direction: misconfigurations cause many costly\Space{
  software and system} failures and severe performance issues.
But, we argue that continued
focus on detection and diagnosis is symptomatic of a more serious
problem: configuration design and implementation are not yet
first-class software engineering endeavors in cloud systems. Little is
known about how and why developers evolve \scdi{}, and the challenges that they face in doing so.

This paper presents a source-code level study of the evolution
of \scdi{} in cloud systems.  Our goal is to understand the
rationale and developer practices for revising initial configuration design/implementation
decisions, especially in response to consequences of
misconfigurations. To this end, we studied 1178 \confcommits{} from
a 2.5 year version-control history of four large-scale,
actively-maintained open-source cloud systems (\hdfs,
\hbase, \spark, and \cassandra). We derive new insights into
the software configuration engineering process.\Space{ We believe
  that} Our results motivate new techniques for \emph{proactively}
reducing misconfigurations by improving the configuration design and
implementation process in cloud systems. We highlight a
number of future research directions\Space{ and opportunities towards
  making configuration design and implementation a key part of
  software engineering}.
\end{abstract}


\blfootnote{This is a preprint of the work (with the same title) published at the 43rd International Conference on
  Software Engineering (ICSE'21). The content should be exactly the same as the
  conference proceeding version, except that the conference version does not
  have the replication package due to its page limit.
  The version is required by ICSE'21 because ``accessibility become vital
  for such a big, distributed and virtual conference''---due to the global COVID-19
  pandemic, ICSE'21 will be a virtual conference in zoomsphere.}

\section{Introduction}
\label{sec:intro}

\begin{table*}[t]
\begin{center}
\footnotesize
\caption{Summary of our findings on \scdi{}, and their implications.}
\begin{tabular}{|l|l|}
\hline
\\[-1em]
\rowcolor{VeryLightGray}\header{Findings about Configuration Interface}  & \header{Implications}         \\
\\[-1em]
\hline
{\bf F.1\ }Software developers often parameterize constant values into  &  {\bf I.1\ }Configuration auto-tuning techniques that consider reliability and func-\\
 configuration parameters. Performance and reliability tuning are common     &     tionality are needed, in addition to performance-only optimization. Timing \\
 rationales for such parameterization.                                        &     parameters are an example (critical to both performance and reliability).  \\
\hline
{\bf F.2\ }Over 50\% of parameterizations were driven by severe consequences        & {\bf I.2\ }Techniques for identifying pathological configuration use cases through \\
  of deficiencies in constant values. Unfortunately, use cases that drove the       & testing and analysis are desired. Tools that can identify and categorize use \\
   parameterization were often poorly discussed or documented.                      & cases could help proactively parameterize deficient constants.        \\
\hline
{\bf F.3\ }Only 28.1\% of default-value changes mentioned systematic testing;      & {\bf I.3\ }Many default values in existing software systems may not be optimal.\\
31.3\% of default changes chose values that work around reported issues.           & Research on how to better select default values is needed. \\
\hline
\hline
\\[-1em]
\rowcolor{VeryLightGray}\header{Findings about Configuration Usage}  & \header{Implications}         \\
\\[-1em]
\hline
{\bf F.4\ }Most configuration-checking code were added as afterthoughts,                & {\bf I.4\ }Proactive parameter value checking and validation can prevent many \\
 postmortem to system failures and performance issues in production.                     & severe consequences (but they are still not a common engineering practice). \\
\hline
{\bf F.5\ }Over 50\% of checks added as afterthoughts are basic (non-emptiness          &  {\bf I.5\ }Automated solutions for generating basic checking code and applying  \\
 and value-range checks); other commits invoked checking code earlier.                   &  them in program early execution phases are useful and feasible. \\
\hline
{\bf F.6\ }Throwing exceptions is common for handling misconfigurations;                & {\bf I.6\ }Automatically correcting configuration errors is feasible and should be \\
 auto-correction is not, missing opportunities to help users handle errors.              & explored in future research. \\
\hline
{\bf F.7\ }Developers often enhance configuration-related log/exception mess-                & {\bf I.7\ }Techniques on automated enhancement of configuration-related log and  \\
   ages by including related parameters and providing guidance.                      &  exception messages to improve misconfiguration diagnosis are needed. \\
\hline
{\bf F.8\ }Reusing existing parameters in different program locations is a com-            & {\bf I.8\ }Tools are needed for identifying and fixing various inconsistencies \\
  mon practice. However, parameter reuse leads to various inconsistencies.                 &  among configuration parameters and their code implementations.  \\
\hline
\hline
\\[-1em]
\rowcolor{VeryLightGray}\header{Findings about Configuration Documentation}  & \header{Implications}        \\
\\[-1em]
\hline
{\bf F.9\ }Inadequate and outdated information are major reasons behind               & {\bf I.9\ }Enforcing complete, up-to-date documentation of configuration info-\\
  the changes that enhance configuration documents.                                            &  mation is still a challenge (despite a lot of research effort). \\
\hline
 {\bf F.10\ }Configuration use cases, parameter constraints and dependencies                  & {\bf I.10\ }Configuration documentation should be systematically augmented \\
 between parameters are commonly added to documents.                             & to include critical, user-facing information. \\
\hline
\end{tabular}
\label{tab:findings}
\end{center}
\end{table*}


Software configuration design and implementation have
  significant impact on the functionality, reliability, and performance
  of large-scale cloud systems\Space{, especially
  those operate at the cloud and datacenter scale}.
The idea behind configuration is to expose\Space{ a number of}
  {\it configuration parameters} which enable\Space{
  system users to perform} {\it deployment}-time system customization.
Using different parameter values, system users (e.g., operators, sysadmins,
  and DevOps engineers) can port a software system
  to different environments,
  accommodate different workloads,
  or satisfy new user requirements.
In cloud systems,
  configuration parameters are changed constantly.
For example, at Facebook, configuration changes are
  committed thousands of times a day,
  significantly outpacing source-code changes~\cite{tang:15}.

With the high velocity of configuration changes,
  misconfigurations (in the form of erroneous parameter values) inevitably become
  a major cause of
  system failures,\Space{ and anomalies that lead to} severe service outages, and downtime.
For example, misconfigurations were
  the second largest cause of service-level disruptions
  in one of Google's main production services~\cite{barroso2018}.
Misconfigurations also contribute to 16\% of production incidents at Facebook~\cite{tang:15},
  including
  the worst-ever outage of Facebook and Instagram that occurred in March 2019~\cite{Shieber:2019}.
Similar statistics and incidents were reported in other\Space{ types of}
  systems~\cite{amvro:16,yin:11,kendrick:12,rabkin:13,gunawi:16,maurer:15,
    oppenheimer:03,nagaraja:04}.
\Comment{Many of those incidents are newsworthy, such as
  the recent worst-ever outage of
  Facebook and Instagram in March 2019~\cite{Shieber:2019}.}

Software configurations\Space{ issues} also impose significant
  total cost of ownership on software vendors, who
need to diagnose user-reported
  failures or performance issues caused by misconfigurations.
Vendors may even have to compensate users,
  if the failures lead to outages and downtime.
Software vendors
  also need to support and help users with configuration-related
  questions, e.g., how to find the right parameter(s) and set the right
  value(s)~\cite{xu:15:2}.
Note that system users
  are often not\Space{ software} developers; they may not understand
  implementation details or they may not be able to debug code~\cite{xu:13,xu:chi:17,xu:16:2}.

Unfortunately,
  \scdi{} have been largely overlooked as first-class software engineering endeavors in cloud systems,
  except for few recent studies (Section~\ref{sec:relatedwork}).
\Space{So far, }The\Space{ main research and development} focus has been on
  detecting misconfigurations and diagnosing their consequences\Space{ (e.g.,
  failures and performance anomalies)}~\cite{xu:13,xu:16,attariyan:10,
    attariyan:12,attariyan:08,zhang:13,zhang:14,
    wang:04,wang:03,santo:16,santo:17,dong:15,
    huang:15,Baset:2017,sayagh:17,sun:osdi:20}.
These efforts tremendously improve {\it system}-level
  defenses against misconfigurations, but they do not address the fundamental need for better
  {\it software} configuration design and implementation.
Yet, better configuration design can  effectively reduce
    user difficulties, reduce configuration complexity while maintaining flexibility,
    and proactively reduce misconfigurations~\cite{norman:83a,norman:83b,maxion:05,xu:15:2}.
Also,\Space{ rigorous, systematic} better configuration implementation can help detect and correct
    misconfigurations earlier to prevent failure damage~\cite{xu:16,xu:13}.
\Comment{
A few recent studies have demonstrated that
  better configuration engineering
  could significantly improve the reliability and usability
  of software systems.
As shown in~\cite{xu:16}, many catastrophic failures
  could be prevented if the software is engineered with the
  practice of rigorous configuration
  checking at the initialization phase.
It is also reported that software configuration design
  is often overly complex and error-prone~\cite{xu:15:2};
  a user-centric design can effectively reduce the complexity
  while maintaining the desired flexibility.
}

The understanding of what constitutes software configuration engineering in cloud systems is preliminary in the literature,
compared with other aspects of
  engineering these software systems
  (e.g., software architecture, modeling, API design, and testing) which are well studied\Space{~\cite{}}.
Meanwhile, we observed that developers struggle to design and implement configurations. For example, we found that developers raise many configuration-related concerns and questions---``{\it is the configuration helpful?}''~(\href{https://github.com/apache/spark/pull/22823}{\SPARK-25676}),
  ``{\it can we reuse an existing parameter\Space{ purpose}?}''~(\href{https://issues.apache.org/jira/browse/HDFS-13735}{\HDFS-13735}),
  ``{\it what is a reasonable default value?}''~(\href{https://issues.apache.org/jira/browse/HBASE-19148}{\HBase-19148}).
Furthermore, we found that
  developers frequently revise
  configuration design/implementation decisions,
  usually after observing severe consequences (e.g., failures and performance issues)
  induced by the initial decisions (Sections~\ref{sec:parameterization} and~\ref{sec:removing:parameters}).

\Comment{
\subsection{Contributions}
\label{sec:contribution}
}

This paper presents a\Space{ comprehensive} source-code level study of
the evolution of \scdi{} in cloud systems, towards filling the knowledge gap
  and better understanding the needs that configuration engineering must meet.
Specifically, we study \checked{1178} \confcommits{} spanning 2.5 years (2017.6--2019.12) in four large-scale, widely-used, and actively-maintained open-source cloud systems (\hdfs, \hbase, \spark, and \cassandra).
\Space{The evolution is embodied in version-control {\it commits} related to
  software configuration in the revision history of the four selected projects;
  furthermore, in these four project, }Each commit that we study is associated with a JIRA/GitHub issue or a Pull Request link which provides more context about the change and the discussions among developers.
(Section~\ref{sec:meth} describes our methodology for selecting
  \confcommits{}\Space{ from the entire revision history}).

Our goal is to\Space{ provide} understand\Space{ing of} current
  configuration engineering practices,
  identify developer pain points,
  and highlight future research opportunities\Space{ for\Space{ achieving} more
  rigorous\Space{, principled} \scdi{}}.
\Space{In particular, }We focus on analyzing commits
  that revise or refine
  initial configuration design or implementation decisions,
  instead of commits that add or remove parameters as code evolves.
These revisions or refinements were driven by\Space{ real-world}
  consequences of misconfigurations\Space{ such as failures and performance issues}.
  Our analysis helps to 1)~understand the rationale for the changes,
  2)~learn design lessons and engineering principles\Space{ qualitatively and quantitatively},
  and 3)~motivate future automated solutions that can prevent such consequences.

To systematically analyze \confcommits,
  we propose a taxonomy of \scdi{} changes in cloud systems along three dimensions:
\myitpara{1)~interface:}why and how developers change the configuration interface
  (parameters, default values and constraints)\Space{ over time}.
\myitpara{2)~usage:}how developers change and improve parameter value
    checking, error-handling, and uses.
\myitpara{3)~documentation:}how developers improve configuration\Space{ information in different forms
  of} documentation.

Note that this paper focuses on cloud systems, instead
  of desktop software or mobile apps,
  because misconfiguring cloud systems results in more far-reaching impact\Space{ due to the scale}.
Moreover, we focus on {\it runtime} configurations~\cite{sayagh18}
  whose values can be changed {\it post-deployment} without\Space{ the need to} re-compiling the software.
Runtime configurations fundamentally differ from
  {\it compile-time} configurations such as \CodeIn{\#ifdef}-based
  feature flags~\cite{meinicke:20}.
But runtime and compile-time configurations have similar problems.
So, there are opportunities to extend techniques that solve problems for one to the other.

This paper makes the following contributions:

\begin{itemize}[topsep=.2ex,itemsep=.2ex,leftmargin=1.75em]

\item[\Contrib{}]\mypara{Study and Insights.}We study code changes to
  understand the evolution of configuration design and implementation in cloud systems. We find
  insights that motivate future research on
  reducing misconfigurations in these systems.

\item[\Contrib{}]\mypara{Taxonomy.}We develop a taxonomy of
  cloud system configuration design and implementation evolution.

\item[\Contrib{}]\mypara{Data.}We release our dataset and scripts at
  ``\url{https://github.com/xlab-uiuc/open-cevo}''
  to help followup research (see Appendix B for the replication package).

\end{itemize}

Table~\ref{tab:findings} summarizes our findings and their implications.


\section{Taxonomy}
\label{sec:background}

\begin{figure}
  \centering
  \includegraphics[width=0.495\textwidth]{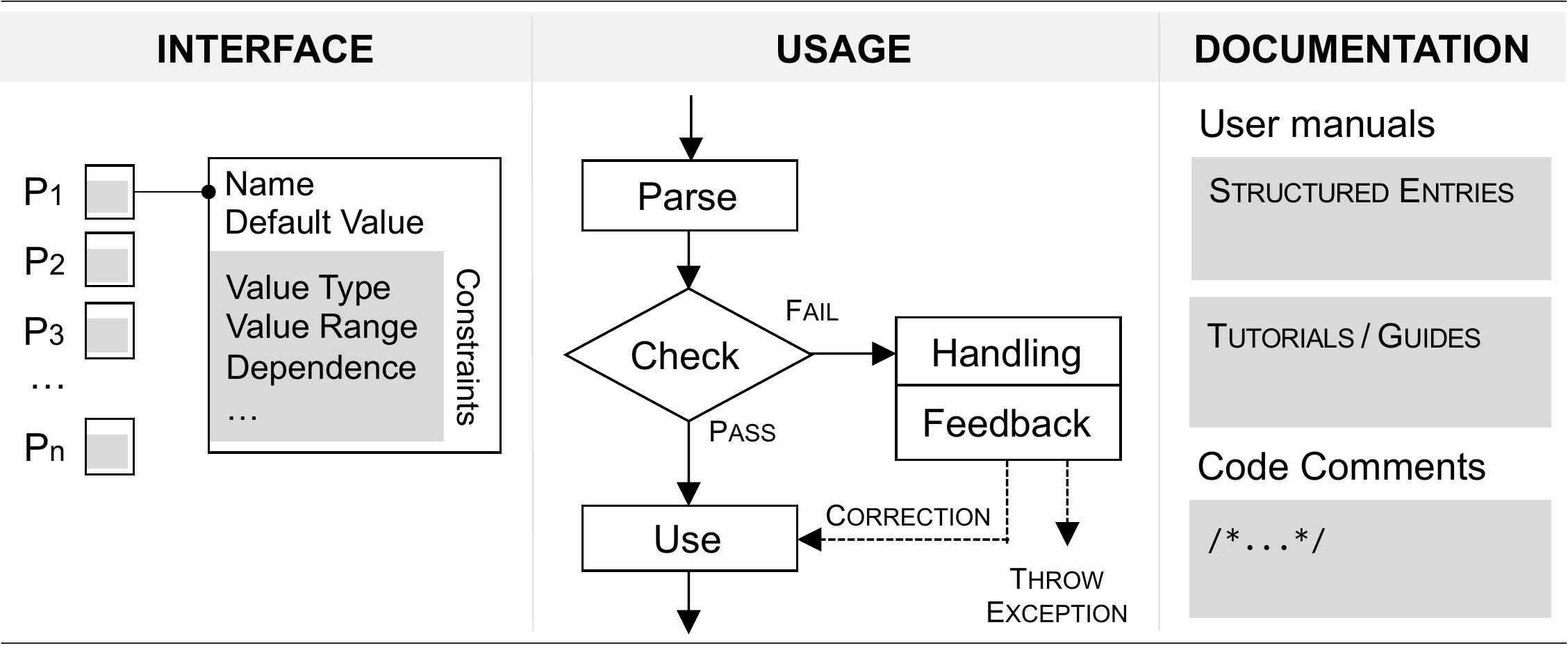}
  \caption{Three parts of our taxonomy of software configuration design and implementation,
    and their components.}
  \label{sec:overview}
\end{figure}

Figure~\ref{sec:overview} shows the three parts of our taxonomy of
 cloud-system configuration engineering: {\it interface}, {\it usage}, and {\it documentation}.
We focus on aspects
  that affect how system users interact
  with configurations, and not on developer-focused aspects
  like variability and testability.
We organize our study along the categories\Space{ of changes in each of
three part, as} shown in Table~\ref{tab:taxonomy}.

\vspace{5pt}
\mypara{Interface.}The configuration interface
    that a system exposes to
    users consists primarily of {\it configuration parameters} (parameters, for short).
  As shown in Fig.~\ref{sec:overview}, a parameter is identified
    by a {\it name} and it typically has a {\it default value}.
  Users can customize system configuration
    by changing parameter values
    in configuration files or by using
    command line interfaces (CLIs).
  Each parameter places {\it constraints} (correctness rules)
    on its values,
    e.g., type, range, dependency on other parameters.
  Values that violate the constraints lead to misconfigurations.
  \Space{As shown }In Table~\ref{tab:taxonomy}, changes that contribute to configuration interface evolution include adding\Space{ new} parameters, removing\Space{ existing}
    parameters, and modifying\Space{ existing} parameters.

\vspace{5pt}
\mypara{Usage.}Fig.~\ref{sec:overview} presents the configuration usage model.
  To use
  a parameter, the software program first
  reads its value from a configuration file or CLI,
   {\it parses} the value and stores it in a program variable.
The variable is then used when the program executes.
In principle, the program {\it checks} the value against the parameter's constraints
 before {\it using} it.
If checks fail, the program needs to {\it handle}
  the error and provide user with {\it feedback messages}.
In Table~\ref{tab:taxonomy}, configuration usage evolution consists of
  changes to all parts of the usage model.

\vspace{5pt}
\mypara{Documentation.}These are natural language descriptions
related to configurations. We consider changes to user manuals and\Space{
  source} code comments---the former are written for system users
while the latter are written for developers.

\begin{table}[]
\footnotesize
\caption{Our taxonomy of configuration engineering evolution.}
\vspace{-2.5pt}
\setlength{\tabcolsep}{2pt}
\begin{tabular}{ll}
\toprule
\rowcolor{VeryLightGray}\multicolumn{2}{c}{\header{INTERFACE (Section~\ref{sec:interface:evolution}})}        \\
  {\bf AddParam}                 & {\bf Add new configuration parameters}   \\
    \quad Add$\rm_{\text{NewCode}}$	        & \quad Add new parameters when introducing new modules    \\
    \quad Add$\rm_{\text{CodeChange}}$       & \quad Add new parameters due to changes in existing code  \\
    \quad Add$\rm_{\text{Parameterization}}$ & \quad Convert constant values to configuration parameters  \\
  {\bf RemoveParam}             & {\bf Remove existing configuration parameters}   \\
    \quad Rmv$\rm_{\text{RmvModule}}$       & \quad Remove parameters when removing existing modules   \\
    \quad Rmv$\rm_{\text{Replace}}$         & \quad Replace parameters with constants or automation  \\
  {\bf ModifyParam}            & {\bf Modify existing configuration parameters}   \\
    \quad Mod$\rm_{\text{Naming}}$		       & \quad Change the name of a configuration parameter    \\
    \quad Mod$\rm_{\text{DefaultValue}}$		 & \quad Change the default value of a configuration parameter  \\
    \quad Mod$\rm_{\text{Constraint}}$	     & \quad Change the constraints of a configuration parameter   \\
\midrule
\rowcolor{VeryLightGray}\multicolumn{2}{c}{\header{USAGE (Section~\ref{sec:behavior:evolution}})}        \\
  {\bf Parse}                 & \textbf{Change configuration parsing code}  \\
  {\bf Check}		    & \textbf{Change configuration checking code}  \\
  {\bf Handle}		    & \textbf{Change configuration error handling code} \\
    \quad Handle$\rm_{\text{Action}}$     & \quad Handling actions (correction and exceptions) \\
    \quad Handle$\rm_{\text{Message}}$    & \quad Feedback messages (log and exception messages) \\
  {\bf Use}                   & \textbf{Change how configuration values are used}  \\
    \quad Use$\rm_{\text{Change}}$        & \quad Change existing code that uses parameters \\
    \quad Use$\rm_{\text{Add}}$           & \quad Add code to reuse a configuration parameter   \\

\midrule
\rowcolor{VeryLightGray}\multicolumn{2}{c}{\header{DOCUMENTATION (Section~\ref{sec:document}})}        \\
  {\bf User manual}          &  \textbf{Change configuration-related user manual content} \\
  {\bf Code comment}         &  \textbf{Change configuration-related source code comments}  \\
\bottomrule
\end{tabular}
\label{tab:taxonomy}
\vspace{15pt}
\end{table}



\section{Study Setup}
\label{sec:meth}

To understand how configuration design/implementation evolve,
  we identified and analyzed {\it \confcommits{}} that
  modify \scdi{}.
Following~\cite{sayagh18}, we refer to the design, implementation, and maintenance of software configuration
  as \emph{configuration engineering}.
We start from commits instead of bug databases (e.g., JIRA and GitHub issues)
  because configuration design and implementation evolution is not limited to bug fixing.\Space{
  Thus, bug databases likely contain less information about configuration evolution than commits.
In fact,}
  All cloud systems that we study\Space{ enforce the practice of} record
  related issue or Pull Request ID(s) in commit messages\Space{, if the commit is part of addressing an
  issue} (Section~\ref{sec:meth:software}).
We found detailed context about changes in the configuration-related commits through developer discussions.
Moreover, commits allow us to analyze the ``{\it diffs}''---the actual changes.


%



\subsection{Target Software and Version Histories}
\label{sec:meth:software}

\Space{In this section, we describe the projects that we study and the portions of their version histories that we use.}
We study configuration design and implementation in four open-source cloud systems, shown in
Table~\ref{tab:software}\Space{ (``\subjectHeader{}'' column)}: \hdfs{},
\hbase{}, \spark{}, and \cassandra{}.\Space{ We selected} These projects\Space{
because they} 1)~have many configuration parameters and
\confcommits{}\Space{ Configuration activities are
  frequent when developing, deploying and using these software.},
2)~are mature, actively-developed and widely-used, with
well-organized GitHub repositories and bug databases,\Space{.
The information is up-to-date and sufficient for us
  to study the evolution of configuration,
3)~use git as their
  version-control system and are hosted on GitHub,} (3)~link
to issue IDs in commit messages\Space{ when applicable},
and 4)~are commonly used subjects in cloud and
datacenter systems research.

In these subjects, we studied \confcommits{} from June 2017 to
December 2019, a 2.5--year time span.
In Table~\ref{tab:software}, ``\header\paramsHeader{}'' is the total number of documented parameters in the
  most recent version, ``\header\allCommitsHeader{}'' is the total number of
commits in the 2.5--year span, and ``\header\studiedCommitsHeader{}'' is the
number of \confcommits{} that we studied. We excluded \confcommits{}
that only added or modified test cases; we expected such commits
to yield less insights on design/implementation evolution.
In total, we
studied \checked{1178} \confcommits{}. \Space{(\checked{1164} contained
issue IDs)}



\begin{table}
\begin{center}
\footnotesize
\caption{Software and their commits that we study.}
\begin{tabular}{llrrr}
\toprule
\header\subjectHeader{}  & \header\descHeader{} & \header\paramsHeader{} & \header\allCommitsHeader{}  & \header\studiedCommitsHeader{}  \\
\midrule
\hdfs{}           & File system     &  \checked{560} & \checked{1618}  & \checked{221}   \\
\hbase{}          & Database        &  \checked{218} & \checked{3516}  & \checked{268}   \\
\spark{}          & Data processing &  \checked{442} & \checked{6194}  & \checked{602}   \\
\cassandra{}      & Database        &  \checked{220} & \checked{1868}  & \checked{87}    \\
\bottomrule
\end{tabular}
\label{tab:software}
\vspace{-10pt}
\end{center}
\end{table}

\Comment{
\begin{table}
\begin{center}
\footnotesize
\caption{Commits collected by different analysis.}
\vspace{-5pt}
\begin{tabular}{ll}
\toprule
Analysis  & Num   \\
\midrule
Commit Messages         & 384      \\
Code Diff         & 794       \\
total & 1178 \\
\bottomrule
\end{tabular}
\label{tab:commit_num}
\vspace{-15pt}
\end{center}
\end{table}
}

\subsection{Data Collection and Analysis}
\label{sec:meth:data_collection}

To find {\it \confcommits{}} within our chosen time span,
  we\Space{ manually analyzed the commit message and the diff\Space{ (both obtained using the \CodeIn{git show} command)}.
We} wrote scripts to automate the analysis of commit messages
  and diffs, filter out irrelevant commits,
  and select likely \confcommits{}.
Then, we manually inspected each resulting commit and its associated issue.
Overall, we collected 384 commits by analyzing commit messages and
  794 commits by analyzing the commit diffs,
  yielding a total of 1178 configuration-related commits.

\subsubsection{Analysis of Commit Messages}
\label{sec:message:analysis}

Keyword search on commit messages is commonly used to find
  related commits\Space{ in a project's version history}, e.g.,~\cite{sai:14,Luo:2014,Bernardo:18,Rigger:19,Hattori:08,dutta2018testing}.
We manually performed a formative study with hundreds of commit messages and found that three strings commonly occur in \confcommits{}: ``config'', ``parameter'' and ``option''.
These strings were previously used in keyword searches\Space{ in previous research}~\cite{sai:14,sayagh18}, and matched 525 times in all four subjects.
We manually inspected these 525 commits and removed commits that did not change configurations, yielding 384 \confcommits.

\Comment{Owolabi: (1)~these numbers do not match Table~\ref{tab:software} (2) It would be good to give one or two broad classes of false positives in the keyword search.}
\Comment{
Why number do not match? This number is just for commit log, we also have commit diff analyze.

Two broad reason:
(1) Most are because ``option'' or ``parameter'' in message which is not configuration option. It is actually easy to tell whether
this option is configuration option from the title and commit log. e.g., ``{\it Add -E option in hdfs ``count'' command to...}'' \href{https://issues.apache.org/jira/browse/HDFS-11647}{\myminus \HDFS-11647}
So basically such FP won't waste much time. However, there are still some commit that just use option/parameter without using config.
So we still keep those two keywords.
(2) Some Commit log mentions ``config'' but the commit is not changing configuration.
e.g., ``{\it Right now spark lets go of executors when they are idle for the 60s (or configurable time) ...}'' \href{https://issues.apache.org/jira/browse/SPARK-21656}{\myminus \SPARK-21656}
}

\subsubsection{Analysis of Commit Diffs}
\label{sec:diff:analysis}
\Space{We observed that searching commit messages by keywords is
insufficient---}Many commit messages do not match during keyword search, even
though the diffs show configuration-related changes.
So, we further analyzed diffs\Space{ (the content that developers changed in each commit)} to
  find more \confcommits{}\Space{ in a more systematic way},
and found additional 794 \confcommits{}\Space{, representing between 1.1$\times$ and 5.4$\times$ more \confcommits{} per project than the keyword-based search (\S\ref{sec:meth:keyword_search})}.
Our diff analysis determines whether diffs
  modify how parameters are defined, loaded, used, or described.
Accurate automated diff analysis requires applying precise taint tracking---treating parameter
  values as initial taints that are propagated along control- and data-flow
  paths~\cite{xu:13,xu:16,rabkin:11,rabkin2:11,attariyan:10,attariyan:12,zhang:13,sai:14,dong:15,Lillack:14}---to
  each commit and its predecessor
  and comparing the taint results in \emph{both} commits.
Such\Space{ precise and} pairwise analysis does not scale well to the 13196 commits in all four projects (Table~\ref{tab:software}).


To scale diff analysis, we used a simple text-based\Space{ approach}
\Space{ which implements the principle of analyzing if a given commit
  changes how configuration is defined, loaded, and used in the
  program. that}search\Space{es Our analysis requires the
  understanding of the} of configuration metadata\Space{ in target
  software projects}, including the 1)~configuration interface
(including how configurations are defined and loaded), 2)~default
configuration file\Space{ structure}, and 3)~message that contains
configuration information.
\Space{configuration }Metadata are expected to be stable in the mature cloud systems
that we study; commits that modify them may yield good insights on configuration evolution.

\Comment{In our experience, these configuration metadata rarely change
  in mature software projects and has been consistent
  in the four software projects used in this study.}

\vspace{3pt}
\mypara{Finding commits that change parameter definitions.}We start
from commits that change default configuration files or parameter
descriptions in those files.  These two locations are key
user-facing parts of configuration design\Space{As configuration is an
  interface for system users}~\cite{xu:13,xu:15}. Thus,\Space{ any}
modification of parameters  (introduction, deprecation, changes to default values, etc.)
likely requires changes to either. This
heuristic was effective: it found 272 additional
\confcommits{}
  with\Space{ a low} an average false positive rate of\Space{ (} 3.2\%\Space{ on average)}.

\vspace{3pt}
\mypara{Finding commits that change parameter loading or setting.}Here, we
leverage knowledge of configuration APIs\Space{ in the subject programs}.  As
reported in prior studies~\cite{xu:16,xu:13,Lillack:14,behrang:15,rabkin:11,rabkin2:11} and validated in
our study, mature software projects have unified, well-defined APIs
for retrieving and assigning parameter values.  For instance, \hdfs{} has\Space{ all
parameter values are retrieved or assigned using}\Space{ a few} getter or setter methods (e.g.,
\texttt{getInt}, \texttt{getBoolean}, declared in a Java class;
each of which has a corresponding setter method (e.g.,
\texttt{setInt}, \texttt{setBoolean}). The other evaluation subjects\Space{
also} follow this pattern.\footnote{This is common in Java and Scala
  projects: the configuration interface typically wraps around
  core library APIs such as \texttt{java.util.Properties}
  to provide configuration getter and setter
  methods.} So, identifying  commits that changed code containing getter or setter method usage
requires a few lines of\Space{ scripting} code using regular expressions. This heuristic\Space{ was
effective: it} found 457 additional
\confcommits{}
with a 19.9\% average false positive rate.

\vspace{3pt}
\mypara{Finding commits that change parameter value data flow.}If a commit changes code
with variables that store parameter values, then that commit is likely related to the data flow of
parameter values.
We implemented a simple text-based taint tracking to track such
variables\Space{. The text-based taint tracking works} as follows.
Once a configuration value is stored in a variable, we add the
variable name to a global taint set.\Space{ The variable name is simply the
left value of an assignment statement, obtained using regular
  expressions.} We perform the tracking for every commit in the
time span that we studied. We do not remove variables from our taint set\Space{ for
completeness}. We output candidate commits where a modified statement
contains a variable name in the taint set. Taint tracking found 31 additional
\confcommits{}
with an average false positive rate of 26.2\%.

\begin{table}[]
  \footnotesize
  \caption{\Confcommits{} by category. Some commits contain changes in multiple
    categories.}
  \begin{tabular}{lccccc}
  \toprule
   & \header{Interface} & \header{Behavior} & \header{Document} & \header{Commit} \\
  \midrule
  \hdfs & \checked{139 (62.9\%)} & \checked{58 (26.2\%)} & \checked{27 (12.2\%)} & \checked{221} \\

  \hbase & \checked{171 (63.8\%)} & \checked{87 (32.5\%)} & \checked{21 (7.8\%)} & \checked{268}  \\

  \spark & \checked{367 (61.0\%)} & \checked{182 (30.2\%)} & \checked{61 (10.1\%)} & \checked{602}     \\

  \cassandra & \checked{54 (62.1\%)} & \checked{32 (36.8\%)} & \checked{5 (5.7\%)}  & \checked{87}     \\
\midrule
  Total & \checked{731 (62.1\%)} & \checked{359 (30.5\%)} & \checked{114 (9.7\%)}  & \checked{1178}  \\
  \bottomrule
  \end{tabular}
  \label{tab:categ_num}
\end{table}

\vspace{3pt}
\mypara{Identifying other \confcommits{}}We applied the same keyword
search on commit messages
(Section~\ref{sec:message:analysis}) to messages that occur in diffs, to capture commits that change
related exception or log messages\Space{. Developers
may change these messages} without modifying any other code.
We found 34 additional
\confcommits{}
with an average false positive rate of 29.2\%.


\subsubsection{Inspection and Categorization}
\label{sec:inspection_cat}

At least two authors independently studied each configuration-related commit and its corresponding issue.
They independently categorized each commit based on the taxonomy\Space{
  presented} in Section~\ref{sec:background},
  and then met to compare their categorization.
When they diverged, a third author was consulted for additional
  discussion until consensus was reached.
Further, in twice-weekly project meetings, the inspectors met with a fourth
  author to review their categorization of 15\% of commits inspected during the week.
  These meetings helped check that understanding of the taxonomy is consistent.
Our experience shows that consistently checking a taxonomy like Figure~\ref{sec:overview}
  with concrete examples significantly improves inter-rater reliability
  and categorization efficiency.

\Space{We would like to}Note that we categorized each commit based on how it revised
    the original configuration design/implementation.
  If a commit adds a new parameter and also a manual entry to document this
    new parameter,
    we treat this commit as Add$_\text{Param}$ (Table~\ref{tab:taxonomy})---the
    commit revises the configuration interface instead of documentation.
   Some commits modify multiple (sub-)parts in our taxonomy.

\Comment{
\blue{
\mypara{Analysis of commits}

We manually analyze and categorize each commit based on the commit log and diff code.
The key principle is to understand how each commit changes the configuration in
different dimensions described in Table~\ref{tab:taxonomy}. Here we illustrate two examples
to show the process:

The variable diskWriteBufferSize is load from configuration by using a configuration getter method.
Then we track how it is used in the commit diff and find out that the value pass to writeBuffer which
is a hard-coded value (1024*1024). Therefore, we categorize it as Parameterization.

Another example is CASSANDRA-14991.
This commit adds error-handling code for using \texttt{server\_encryption\_options} and \texttt{client\_encryption\_options}.
We categorize this commit into "Evolution of Error-handling Code (Section 5.2)."
}

\vspace{3pt}
\mypara{Analysis of JIRA/GitHub Issues}All \checked{1178} \confcommits{}
that we collected\Space{ in this (\S\ref{sec:message:analysis} and
  \S\ref{sec:diff:analysis})} have associated JIRA/GitHub issues
(\checked{1164} commits) or are linked to Pull Requests (\checked{14} commits).
We\Space{ manually} inspected the description and developers discussion
in these issues and PRs, to
obtain more\Space{ detailed background and} context about each commit\Space{s},
which helps us validate the categorization and finding insights.
}

\subsubsection{Data Collection Results}

Table~\ref{tab:categ_num} shows the studied \confcommits{}
along the three parts of our taxonomy\Space{ in Table~\ref{tab:taxonomy}
(interface, behavior, and documentation)}\Space{ across the four
  studied software projects}. There is a significant number of
commits in each part.
The rest of this paper summarizes
our\Space{ per-commit} analysis and provides insights on how \scdi{} evolve
along these three parts.




\section{Configuration Interface Evolution}
\label{sec:interface:evolution}

Changes to the configuration
{\it interface} were the most common\Space{ evolution types},
  compared with behavior or documentation changes (Table~\ref{tab:categ_num}).
\Space{As shown in Table~\ref{tab:categ_num}, interface evolution
  contributes to \red{XX\%--YY\%} of all the changes across the
  studied software projects.} We focus on analyzing changes to \emph{configurability}---the level of user-facing configuration flexibility\Space{
  that the system}---(Section~\ref{sec:configurability})
  and default values (Section~\ref{sec:default_value}).
We omit other kinds of configuration
  interface changes which are often routine and cannot directly lead to misconfigurations.

\subsection{Evolution of User-Facing Configuration}
\label{sec:configurability}

Table~\ref{tab:interface} shows our categorization of changes to configurability.\Space{Most configurability-related} Most changes
add or remove parameters;\Space{ for comparison,}
 per project, removal\Space{ existing parameters} is 5.1$\times$ to
  21.2$\times$ less frequent than addition (with an average of 8.4$\times$).
\Space{Upon further analysis, }We find that adding or removing parameters occur naturally\Space{ consequence of} during software
evolution---\Space{new }parameters are added with new code, and\Space{ parameters}
removed with code deletion.
\Space{In the rest of this section. So,}We do not focus on co-addition or
co-removal of parameters with code.
Rather, we focus on changes\Space{ developers make} that revise
previous configuration engineering decisions by
1)~parameterizing constants and 2)~eliminating parameters or converting them to constants.

\Comment{Owolabi: why should we not study rationales for adding and deleting more deeply in the future?}



Our data corroborates a prior finding~\cite{xu:15:2} that configuration interface complexity
increases rapidly over time, as more parameters are added than are removed.
Complexity measures the size of the configuration space (number
of parameters multiplied by the number of all their possible values).
\Space{Techniques and practices}Approaches for dealing with the rapid growth rate\Space{
interface complexity} are\Space{ urgently} desired.
\Space{Specifically, }Variability modeling~\cite{Lotufo:10,She:10,Damir2019FSE,Berger2013TSE,Liebig:10} which is extensively
  researched for compile-time configurations,
  can potentially be extended to understand and manage runtime configuration complexity.

\subsubsection{Parameterization}
\label{sec:parameterization}
\Space{We observe that software }Developers often\Space{ parameterize constants---they} convert constants into
parameters\Space{---} after discovering that one constant cannot satisfy all use cases. We find \checked{142} commits
that parameterize  \checked{169} constants (\checked{169} parameterizations). We report on
1)~rationales for the parameterizations,
2)~how developers identify constants to parameterize,
3)~use cases that made constants insufficient,
and 4)~how developers\Space{ currently use to} balance increase in configuration complexity
(caused by adding new parameters) with the need for flexibility (which necessitates parameterization).
Our results have\Space{ significant} ramifications for
configuration interface design:
we provide understanding\Space{ necessary} for managing the
configurability versus simplicity tradeoff.
The rationales for parameterization also\Space{ has implications that} motivate\Space{ the need for}
configuration parameter auto-tuning.

\vspace{5pt}
\mypara{\bf Rationales for parameterization.}These
 include: performance
tuning, reliability, environment setup, manageability, debugging,
compatibility, testability, and security.
Table~\ref{tab:purpose_constant2config} shows, for each rationale, the number of commits\Space{ (\header{\#Commits})} and parameters\Space{ (\header{\#Params})}\Space{ that
  we encountered}, an example parameter, and a description.
We discuss the top two rationales, due to space limits.
Performance tuning was the top rationale for parameterizing constants,
  involving \checked{39.6\%
  (67/169)} of parameters in \checked{56}\Space{ Add$\rm_{const2config}$} commits\Space{ in
Table~\ref{tab:purpose_constant2config}}.\Space{. Parameterizing for performance tuning makes sense:}
Different workloads need different values, so it is hard to
find one-size-fits-all constants.\Space{ So, developers parameterize
constants\Space{ that are then exposed} to allow users
customize\Space{ those parameters} for their workloads.}
\Space{Among performance-tuning-related parameterization,}
%
%
Resource-related (e.g., buffer size and thread number),
  feature selection (turning on/off features with performance impact, e.g.,
  monitoring),
  and timing logic (mostly timeouts and intervals) were the main resulting parameter types, with
  20.9\% (14/67), 37.3\% (25/67), and 14.9\% (10/67) new parameters, respectively.
Others 26.9\% (18/67) set algorithm-specific parameters (e.g., weights and sample sizes).

\begin{table}[]
  \small
  \centering
  \setlength{\tabcolsep}{4pt}
  \caption{Statistics on configuration interface changes.}
  \scalebox{0.9}{
  \begin{tabular}{lccccc}
  \toprule
   & \header\hdfs & \header\hbase & \header\spark & \header\cassandra & \header{Total} \\
  \midrule
  {\bf AddParam}             			& \checked{106} & \checked{122} & \checked{277} & \checked{42} & \checked{547}  \\
      \quad Add$_\text{NewCode}$		& \checked{54} & \checked{55}  & \checked{143} & \checked{23} & \checked{275}  \\
      \quad Add$_\text{CodeChange}$             	 & \checked{16} & \checked{34} & \checked{72} & \checked{8} & \checked{130} \\
      \quad Add$_\text{Parameterization}$   	  & \checked{36} & \checked{33} & \checked{62} & \checked{11}  &  \checked{142} \\
  \midrule
  {\bf RemoveParam}                                 	& \checked{5} & \checked{24} & \checked{30} & \checked{6} & \checked{65}  \\
      \quad Rmv$_\text{RmvModule}$    		  & \checked{3} & \checked{16} & \checked{25} & \checked{4} & \checked{48} \\
      \quad Rmv$_\text{Replace}$    		& \checked{2} & \checked{8}  & \checked{5} & \checked{2}  & \checked{17} \\
  \midrule
  {\bf ModifyParam}                                 & \checked{28} & \checked{25} &  \checked{60} & \checked{6} & \checked{119} \\
      \quad Mod$_\text{Naming}$		& \checked{5}  & \checked{8} & \checked{30} & \checked{1} & \checked{44} \\
      \quad Mod$_\text{DefaultValue}$		& \checked{19} & \checked{14} & \checked{20} & \checked{3} & \checked{56}  \\
      \quad Mod$_\text{Constraint}$	& \checked{4} & \checked{3} & \checked{10} & \checked{2} & \checked{19} \\
  \bottomrule
  \end{tabular}
  }
  \label{tab:interface}
\end{table}

\begin{table*}[t]
\footnotesize
\setlength{\tabcolsep}{2pt}
\caption{Statistics and examples of developers' rationales for parameterization (excluding two commits that lacks information). }
\scalebox{0.95}{
\begin{tabular}{lccll}
\toprule
\header{Rationale}  & \header{\#Commit} & \header{\#Param} & \header{Example New Parameter}    & \header{Limitation of previous constant}         \\
\toprule
Performance  & 56 & 67   &  {\scriptsize spark.sql.codegen.cache.maxEntries}   & The cache size does not work for online stream processing~(\href{https://issues.apache.org/jira/browse/SPARK-24727}{\SPARK-24727})\\
\midrule
Reliability    & 28 & 37 &  {\scriptsize spark.sql.broadcastExchange.maxThreadThreshold} & Out of memory if thread-object garbage collection is too slow~(\href{https://issues.apache.org/jira/browse/SPARK-26601}{\SPARK-26601})  \\
\midrule
Manageability   & 20 & 20 & {\scriptsize dfs.federation.router.default.nameservice.enable} & Enable the default name service to store files\Space{ at specified locations}~(\href{https://issues.apache.org/jira/browse/HDFS-13857}{\HDFS-13857})  \\
\midrule
Debugging & 8 & 9  & {\scriptsize spark.kubernetes.deleteExecutors}  & Disable auto-deletion of pods for debugging and diagnosis~(\href{https://issues.apache.org/jira/browse/SPARK-25515}{\SPARK-25515}) \\
\midrule
Environment  & 8 & 13  & {\scriptsize dfs.cblock.iscsi.advertised.ip}    & \Space{An iSCSI}Allows server and target addresses to be different~(\href{https://issues.apache.org/jira/browse/HDFS-13018}{\HDFS-13018}) \\
\midrule
Compatibility  & 13 & 13 & {\scriptsize spark.network.remoteReadNioBufferConversion}  &  Add the parameter to fall back to an old code path~(\href{https://issues.apache.org/jira/browse/SPARK-24307}{\SPARK-24307}) \\
\midrule
Testability  & 3 & 4   & {\scriptsize spark.security.credentials.renewalRatio}   &  May not need to be set in production but can make testing easier~(\href{https://github.com/apache/spark/pull/20657}{\SPARK-23361}) \\
\midrule
Security     & 4 & 4   & {\scriptsize spark.sql.redaction.string.regex}  & The output of query explanation can contain sensitive information~(\href{https://issues.apache.org/jira/browse/SPARK-22791}{\SPARK-22791}) \\
\bottomrule
\end{tabular}
}
\label{tab:purpose_constant2config}
\end{table*}

Reliability, with\Space{ \checked{21.9\%} (} \checked{37} of \checked{169} of the parameterizations, was the second most
common rationale.
\Space{\checked{45.9\% (17/37)}}Of these, \checked{17} were caused by hardcoded timeout values
  that led to constant request failures in the reported deployments,
	so developers made them configurable.
Note that new timing parameters were created for
  both reliability and performance tuning.
For example, in \hdfs{}, a new timing\Space{-related} parameter\Space{,
  \CodeIn{\seqsplit{regionserver.hfilecleaner.thread.timeout.msec}}}
  was created to improve performance. The previous constant was causing a ``{\it \Space{
  While deleting a file never complete (strange but possible),
  then }delete file task\Space{ needs} to wait for\Space{ a max
  of 60 seconds which might be}... too long}''~(\href{https://issues.apache.org/jira/browse/HBASE-20401}{\HBase-20401}).
  But, another \hdfs{} timing\Space{-related} parameter\Space{, \CodeIn{\seqsplit{cblock.rpc.timeout.seconds}},} was created to improve reliability. The previous constant\Space{value}
  	{was too small,}
  \Space{of 300s}causing ``{\it \Space{Volume creation }timeouts while creating 3TB
  volume\Space{ because of too many containers}}''~(\href{https://issues.apache.org/jira/browse/HDFS-12210}{\HDFS-12210}).

\begin{shadedbox}
  {\bf \header{Discussion:}} Configuration auto-tuning techniques
    that consider reliability and functionality are needed, in addition to
  performance-only optimization~\cite{aken:17,
    wang:18, zheng:07, yu:18, nair:17, siegmund:15,
    Nair:2018, Hsu:2018, duan:09, zhu:17,
    Jamshidi:17, Jamshidi:18, xi:04, ye:03, Herodotou:2011, osogami:06,
    Krishna:19, hoffmann:11, Herodotou:11}.
Specifically, timing\Space{ related} parameters have important implications to both reliability
  and performance; however, not much work has been done on auto-tuning
  timing parameters (e.g., timeouts and intervals).
\end{shadedbox}


\Comment{@Owolabi: the point I wanted to make is that there are multiple
  objectives for setting configuration values --- it seems to me that
  all the existing auto-tuning techniques only focus on performance.}



\vspace{8pt}
\mypara{\bf How developers find constants to parameterize.}
\checked{54.4\% (92/169)} of parameterizations\Space{ (\red{56.3\% (63/112)} of commits),}
  were {\it postmortem} to
  severe consequences\Space{ in the field}, e.g., system failures, performance
  degradation, resource overuse, and incorrect results.
Among previous constants for these,
\checked{40.2\% (37/92)}\Space{ of the constants} led to performance
degradation\Space{ such as \red{prolonged execution time}, \red{excessive
  end-to-end latency}, and \red{low throughput}};
\checked{35.9\% (33/92)}\Space{ of the constants} caused
severe failures\Space{ such as job-level and system-level failures};
\checked{19.6\% (18/92)}\Space{ of the constants}
led to incorrect or unexpected results (e.g., data
loss and wrong output);
and \checked{4.3\% (4/92)}\Space{ of the constants} resulted in
resource overuse.



\Comment{
\vspace{5pt}
{\color{gray}
\mypara{\bf \header{Implications}}Over 50\% of parameterizations were
found deficient---not satisfying all use cases---after
negative consequences ensued.\Comment{may not be "after failure". I guess is "after bad consequence" }
Future work should develop \emph{proactive} techniques for
detecting\Space{ and warning developers about} deficient constants
and automate parameterization; the latter could\Space{ also} assist\Space{ with}
performance testing of cloud and datacenter software. \Comment{Any other deep stuff?}
}
}

\begin{shadedbox}
{\bf \header{Discussion:}}
Despite the efforts in parameterization,
  developers still overlook deficient constants that may lead to severe consequences (e.g., failures and performance issues).
\textit{Proactive} techniques for detecting deficient constants and for automating
  parameterization are needed; the latter could\Space{ also} assist\Space{ with}
  performance testing of cloud systems.
\end{shadedbox}


\vspace{5pt}
\mypara{\bf Describing use cases that prompt parameterization.}\Space{As discussed above, The main reason for introducing
parameters is that a constant does not satisfy
all use cases and workloads. Thus, to improve understanding of the
resulting configuration design, }Use cases where constants
were deficient should be described fully\Space{ (if not defined)} to help users set
correct values.
But,\Space{ we find that} developers described the {\it concrete}
use cases that prompt parameterization for only
\checked{37.9\% (64/169) parameters}. Others discussed use cases
either {\it vaguely} \checked{(45.0\% or 76/169 parameters)} or provided \emph{no information}
\checked{(17.1\% or 29/169 parameters)}\Space{, thus failing to provide guidance
  for improving the understanding of
resulting parameters}. Table~\ref{tab:des_level} shows examples.

\Comment{
\tianyin{@Owolabi: One potential direction is to automatically identify the
  scenarios and apply the corresponding configuration values.}
}

\begin{shadedbox}
{\bf \header{Discussion:}} Future work should
identify and document use cases and workloads,
  including which parameters can be tuned, and suggest beneficial configuration
  values that are designed for concrete use cases.
\end{shadedbox}

\begin{table}
\footnotesize
\centering
\setlength{\tabcolsep}{0pt}
\caption{Use-case description of parameterization changes.}
\scalebox{0.85}{
\begin{tabular}{ll}
\toprule
\header{Level} & $\ \ \ $\header{Example} \\
\toprule
Concrete & $\ $``{\it \scriptsize Volume creation times out while creating \textbf{3TB} volume}''~(\href{https://issues.apache.org/jira/browse/HDFS-12210}{\HDFS-12210}). \\
\midrule
Vague    & ``{\it \scriptsize If \textbf{many} regions on a RegionServer, the default will be not enough}''~(\href{https://issues.apache.org/jira/browse/HBASE-21764}{\HBase-21764}). \\
\midrule
No Info  &  ``{\it \scriptsize It \textbf{would be better} if the user has the option instead of a constant}''~(\href{https://issues.apache.org/jira/browse/SPARK-25233}{\SPARK-25233}). \\
\bottomrule
\end{tabular}
}
\label{tab:des_level}
\end{table}

\vspace{5pt}
\mypara{\bf Balancing flexibility and simplicity}Configuration interface design must balance flexibility
(i.e., configurability) with simplicity~\cite{xu:15:2}. New
parameters increase flexibility (by handling additional use cases),
but increase interface complexity (thus reducing usability).
\checked{16.2\% (23/142)} of parameterization commits\Space{issue discussion, PR discussion} contained developer discussions on the
flexibility-simplicity tradeoff.  Most\Space{ of them} discussed\Space{
the} estimated prevalence of use cases for parameterization---it
is not worth increasing interface complexity
for rare\Space{ly-occurring} use cases---and typically involve\Space{ one or
two} advanced users, e.g., ``{\it
  admittedly, this...\Space{ [\texttt{.blockWhenSyncLagsNanos}]}is\Space{ only} an
  expert-level setting,\Space{ but} useful in some cases}''~(\href{https://issues.apache.org/jira/browse/CASSANDRA-14580}{\Cassandra-14580}).
We also found\Space{ subjective developer discussions and} developers' debates on whether\Space{
  a constant should be} to parameterize
  (\href{https://issues.apache.org/jira/browse/CASSANDRA-12526}{\Cassandra-12526},
   \href{https://issues.apache.org/jira/browse/HDFS-12496}{\HDFS-12496},
   \href{https://github.com/apache/spark/pull/23090}{\SPARK-26118}).

A {\it middle-ground} solution\Space{ in these discussions} is to
parameterize without documenting or exposing the parameter, e.g.,
``{\it although...\Space{ maybe}not widely
  used, I could see allowing control\Space{ of this}...via an undocumented
  parameter}''~(\href{https://github.com/apache/spark/pull/21433}{\SPARK-23820}).
With this practice, not all but the most advanced users know of such parameters.
We found that
\checked{58.0\% (98/169)} of the
  newly added parameters were not documented in the parameterization commit, indicating
  that these parameters were first added as middle-ground solutions.

\begin{shadedbox}
{\bf \header{Discussion:}} Further studies are needed on 1) if and why
  undocumented parameters are eventually documented, and 2) how often and why  (expert) users
  modify un-exposed parameters, in order to understand the intent and utility of such parameters.
\end{shadedbox}

Specifically, visibility conditions from variability modeling~\cite{Damir2019FSE,Berger2013TSE,Berger:10}
  can be extended to manage the tradeoff of flexibility versus simplicity,
  which can benefit navigation support and user guidance~\cite{barrett:04,takayama:06,haber:07,jin:14}.
Currently, visibility conditions are mainly designed for Boolean feature flags based
  on dependency specifications (e.g., in {CDL and KConfig);
  complexity metrics and variability analysis for other parameter types (e.g., numeric and strings)
  are needed.




\subsubsection{Removing Parameters}
\label{sec:removing:parameters}

\Space{Parameter removal
is infrequent relative to addition (Table~\ref{tab:interface}).} Understanding
parameter removal can yield insights on reducing configuration
interface complexity~\cite{xu:15:2,meinicke:20}.
We examined all \checked{17} \confcommits{} that removed a parameter (not co-removal with code).
All \checked{17} removed parameters were
converted to constants or code logic was added that obviated them.
\checked{14}\Space{ of \checked{17}} removed parameters were converted to constants: 11 to their default values
and 3 to safe values.
Developers
mentioned that \checked{8} of the \checked{17} parameters had no clear use case
  (e.g.,~\href{https://issues.apache.org/jira/browse/HBASE-8518}{\HBase-8518},
   \href{https://issues.apache.org/jira/browse/HBASE-18786}{\HBase-18786}),
   or required users to understand implementation details (e.g.,~\href{https://issues.apache.org/jira/browse/CASSANDRA-14108}{\Cassandra-14108}).
\checked{Three} parameters\Space{ even} confused users or might lead to severe
  errors~(e.g.,~\href{https://issues.apache.org/jira/browse/SPARK-26362}{\SPARK-26362}).
%
%







\checked{Three} of \checked{17} removed parameters were obviated by new automation logic.
For example, in \href{https://issues.apache.org/jira/browse/HBASE-21228}{\HBase-21228},
\CodeIn{\seqsplit{hbase.regionserver.handler.count}} which
specified the number of concurrently updating threads to be garbage
collected in a Java \CodeIn{ConcurrentHashMap}, was removed after
developers switched to
\CodeIn{\seqsplit{ThreadLocal<SyncFuture>}} which automatically garbage collects terminated threads.
This example shows how implementation choices could affect configuration complexity.

\begin{shadedbox}
{\bf \header{Discussion:}} Future studies can evaluate the utility and impact
  of each parameter (e.g., by analyzing if and how often deployed
  values are equal or similar to the default values).
  Configurations with low utility can be replaced with constants (e.g., default values).
\end{shadedbox}

\Comment{@Owolabi: is this an implication: basically adding a configuration
  parameter should be a metric to consider in the implementation.}

\Comment{@Owolabi: perhaps we can make a point that
  those two principles can be generalized to further eliminate and
  simplify existing configuration design. Owolabi: that would require
  a way to first know the use cases under which values of all
  parameters are useful; I don't think that's feasible}

\subsection{Evolution of Default Values}
\label{sec:default_value}

Parameter default values are important to the usability of\Space{ highly}
configurable systems; they provide users with good
starting points for setting parameters without needing to understand the
entire configuration space.  Thus,\Space{ to minimize user configuration
efforts,} developers usually choose default values that satisfy
common use cases. Ideally, a default value applies under most common
workloads, without causing failures\Space{ and does not lead to software failure (including
performance degradation) in any workload}
	(\href{https://issues.apache.org/jira/browse/HBASE-16417}{\HBase-16417},
	\href{https://issues.apache.org/jira/browse/HBASE-20390}{\HBase-20390},
	\href{https://issues.apache.org/jira/browse/HDFS-11998}{\HDFS-11998}).
The ``Mod$\rm_{DefaultValue}$'' row in Table~\ref{tab:interface} shows\Space{ that
there were} \checked{56} commits that changed \checked{81} default\Space{ parameter}
values\Space{ in all four subjects}\Space{ during the two-year range of
  \confcommits{} that we studied, 2--19 default values are changed
  across the studied modules}. We discuss\Space{ here the reasons} why default
values changed and how new default values were chosen.


\Comment{Tianyin:This is a very deep question but the answer is not
  answering the question at all. Let me elaborate the question. --- as
  we discussed over Skype, the simplest way is to focus on why the old
  default starts to break things and need the new values --- something
  has to be changed, either the workload, or the hardware, or the
  scale, or the environment.}

\vspace{3pt}
\mypara{Reasons for changing default values.}
  We observe proactive and reactive default value changes.
  \checked{38.3\% (31/81)} default-value changes were proactive, including
	  1) enabling a previously disabled feature flag (\checked{32.3\% (10/31)}),
	   e.g., ``{\it running the feature in production for a while with no issues,
     so enabled the feature by default}'' (\href{https://issues.apache.org/jira/browse/HDFS-7964}{HDFS-7964}),
	  2) performance reasons (\checked{35.4\%, 11/31}), e.g.,
    ``{\it sets properties at values yielding optimal performance}'' (\href{https://issues.apache.org/jira/browse/HBASE-16417}{\HBase-16417}), and
	  3) supporting new use cases (\checked{32.3\%, 10/31}), e.g., {\it ``it may be a common use case to ...list
	  queries on these values}'' (\href{https://issues.apache.org/jira/browse/Cassandra-14498}{\Cassandra-14498}).

The remaining \checked{61.7\% (50/81)} of default value changes were reactive to
  user-reported issues, including
    1) system failures and performance anomalies due to not supporting
      new workloads, deployment scale, hardware, etc (\checked{50.0\%, 25/50}),
    2) inconsistencies with user manual (\checked{38.0\%, 19/50}),
    and 3) working around software bugs (\checked{12.0\%, 6/50}), e.g.,
    {\it ``we set the parameter to false by default for Spark 2.3 and re-enable
    	  it after addressing the lock congestion issue''} (\href{https://issues.apache.org/jira/browse/SPARK-23310}{\SPARK-23310}).

\vspace{3pt}
\mypara{Choosing new values.}It is straightforward to change
new default values for
Boolean and enumerative parameters, given their small value ranges.
So, we describe how new default
  values of \checked{32} {\it numeric} parameters were chosen (excluding
  those that fix default value inconsistency (e.g., \href{https://issues.apache.org/jira/browse/HBASE-18662}{\HBase-18662}).
Only \checked{28.1\% (9/32)} numeric parameters had systematic
performance testing and benchmarking mentioned in the JIRA/GitHub
issues\Space{ accompanying the default-value change}.
Later commits reset these new default values,
despite the initial testing and benchmarking. For example, \hbase{}
developers\Space{ said they had} performed ``{\it write-only workload
  evaluation...read performance in read-write workloads. We
  investigate several settings...\Space{ of hardware (SSD, HDD), key
    distribution (Zipf, uniform), with multiple settings\Space{ of the
    system, and compare measures like write throughput, read latency,
    write volume, total gc time, etc.}}}''~(\href{https://issues.apache.org/jira/browse/HBASE-16417}{\HBase-16417}). Yet,
we found three later commits that changed the default value of the
same parameter to different numbers.
For \checked{31.3\% (10/32)} of numeric parameters, new default values were
chosen by adjusting the previous default values to resolve production
failures. In many of these cases, usually without high
confidence in the new default values, developers simply chose
values that resolve the\Space{ reported} problem(s).  Examples: ``{\it It
  probably makes sense to set it to something\Space{ even} lower\Space{
  still}}''~(\href{https://issues.apache.org/jira/browse/SPARK-24297}{\SPARK-24297}), or ``{\it I'm thinking something like
  3000 or 5000 would be\Space{ a} safer\Space{ bet.}}''~(\href{https://issues.apache.org/jira/browse/HBASE-18023}{\HBase-18023}).
We found no\Space{ concrete} information on\Space{ how new default values
were selected for} the remaining \checked{40.6\% (13/32)} numeric
parameters.

We observe that backward compatibility and safety are common considerations
  in selecting new default values. New default values that radically
  change system behavior are often considered inappropriate
  (e.g., \href{https://issues.apache.org/jira/browse/HBASE-18662}{\HBase-18662}).

\begin{shadedbox}
{\bf \header{Discussion:}} \Space{In the studied cloud systems,
  }Default value changes are often reactive to the reported issues,
  without systematic assessment.
Systematic testing and evaluation of
  new (and\Space{ even} existing)
  default values are needed.
\end{shadedbox}

Dynamic workloads and heterogenous deployments necessitate continuous and incremental changes to default values.
Future work could maintain a set of
  default values (instead of one) for\Space{ various} typical
  workloads, hardware, and scale.

\subsection{Summary}

There is an unmet need for {\it practical} configuration
  automation techniques and tools for
choosing and testing parameter values---why do cloud system developers still change parameter values statically
rather than using parameter automation?
  There is also need
for automatic ways of identifying workloads or use cases for which
default values (and even constants) are ill-suited.
Such automatic workload identification
approaches can help developers to better 1)~decide which constant
values need to be parameterized, 2)~understand when their
current default values will lead to system failures,
and 3)~come up with better tests and benchmarks for default values.

\section{Configuration Usage Evolution}
\label{sec:behavior:evolution}

\begin{table}
\footnotesize
\setlength{\tabcolsep}{4pt}
\caption{Statistics on configuration usage evolution.}
\begin{tabular}{lccccc}
\toprule
 & \header\hdfs & \header\hbase & \header\spark & \header\cassandra & \header{total} \\
\midrule
 {\bf Parse}             			& \checked{5} & \checked{14} & \checked{59} & \checked{7} & \checked{85} \\
 \midrule
  {\bf Check}            	& \checked{7} & \checked{20} & \checked{29} & \checked{11} & \checked{67} \\
 \midrule
  {\bf Handle}                 & \checked{12} & \checked{18} & \checked{20} & \checked{2} & \checked{52}  \\
     \quad Handle$_\text{Action}$	        	& \checked{8} & \checked{6} & \checked{4} & \checked{1} & \checked{19}  \\
     \quad Handle$_\text{Message}$		 & \checked{4} & \checked{12} & \checked{16} & \checked{1} &  \checked{33} \\
 \midrule
   {\bf Use}             		& \checked{34} & \checked{35} & \checked{74} & \checked{12} & \checked{155} \\
     \quad Use$_\text{Change}$		& \checked{7}  & \checked{10} & \checked{25} & \checked{3} &  \checked{45} \\
     \quad Use$_\text{Add}$	& \checked{27} & \checked{25} & \checked{49} & \checked{9} & \checked{110} \\
\bottomrule
\end{tabular}
\label{tab:behavior}
\end{table}

\begin{table*}
\begin{center}
\footnotesize
\centering
\caption{Examples of consequences that can be prevented by adding configuration checking code.}
\begin{tabular}{lll}
\toprule
\header{Consequence}    & \header{Example Parameter}    & \header{Description}       \\
\toprule
Runtime Error     & {\scriptsize hbase.bucketcache.bucket.sizes} &  If value is not aligned with 256, instantiating a bucket cache throws IOException~(\href{https://issues.apache.org/jira/browse/HBASE-16993}{\HBase-16993})  \\
\midrule
Early Termination                  & {\scriptsize commitlog\_segment\_size\_in\_mb}                      &   If value $\ge$ 2048, Cassandra throws an exception when creating commit logs~(\href{https://issues.apache.org/jira/browse/CASSANDRA-13565}{\Cassandra-13565})   \\
\midrule
Service unavailability             & {\scriptsize spark.dynamicAllocation.enabled}   &  Running barrier stage with dynamic resource allocation may cause deadlocks~(\href{https://issues.apache.org/jira/browse/SPARK-24954}{\SPARK-24954})  \\
\midrule
Unexpected Results                 & {\scriptsize spark.sql.shuffle.partitions} &   If the value is 0, the result of a table join will be an empty table~(\href{https://issues.apache.org/jira/browse/SPARK-24783}{\SPARK-24783})    \\
\bottomrule
\end{tabular}
\label{tab:conseq_check}
\end{center}
\end{table*}

We present results on configuration usage evolution (recall the configuration usage model described in Fig.~\ref{sec:overview} and Section~\ref{sec:background}).
\Comment{To use a configuration parameter, the software program
  {\it parses} its value and stored it in a program variable.
 The value is then {\it checked} before {\it using} it during
 execution. If checks fail, the program {\it handles}
 the configuration error and provides user {\it feedback} by either (1)~throwing an exception,
 or (2)~logging and correcting the error then resuming execution.}
Across the four cloud systems,
  \checked{26.2\%--36.8\%} of commits changed parameter usage (Table~\ref{tab:behavior}).
We describe changes to checking, error handing, and
  use of parameters.
We omit changes to parsing APIs (e.g.,~\href{https://github.com/apache/spark/commit/dad2d826}{\SPARK-23207}).


\subsection{Evolution of Parameter Checking Code}
\label{sec:usage:check}

Proactively checking parameter values is key to preventing\Space{ runtime failures
  and performance anomalies caused by} misconfigurations~\cite{xu:16}.
However, we find that many parameters had no \emph{checking code} when they were introduced.
  Checking code was added {\it reactively}\Space{ in the aftermath}:
1)~\checked{74.6\% (50/67)} of commits that changed checking code occurred after users reported
    runtime failures, service unavailability,
    incorrect/unexpected results, startup failures, etc.
    (Table~\ref{tab:conseq_check} shows examples).
2)~\checked{In 14.9\% (10/67)} commits, developers {\it proactively}
  added or improved the checking code; 2 of them\Space{ added checks
  for six parameters by} applied reactively-added
  checking code to other parameters with similar types (e.g.,~\href{https://issues.apache.org/jira/browse/CASSANDRA-13622}{\CASSANDRA-13622}).
3)~We did not find sufficient information of the other \checked{7 commits}.

\begin{figure}
\begin{minipage}{0.97\linewidth}
  \centerline{
\begin{lstlisting}[language=diff]
+ if (writeTables==null || writeTables.isEmpty()) {
+   throw new IllegalArgumentException(
+     "Configurtion parameter" +
+     OUTPUT_TABLE_NAME_CONF_KEY + "cannot be empty")}
\end{lstlisting}}
  \vspace{3.5pt}
  \centerline{\small (a) Add a \CodeIn{NOT-NULL} check (\href{https://github.com/apache/hbase/commit/293cb87d52d6ccc98f0d03387f9dc07dc4522042}{\HBASE-18161})}
  \vspace{3.5pt}
\end{minipage}
\begin{minipage}{0.97\linewidth}
  \centerline{
\begin{lstlisting}[language=diff]
+  if (bucketSize % 256 != 0) {
+    throw new IllegalArgumentException(  
+      "Illegal value:" + bucketSize +  
+      "configured for" + BUCKET_CACHE_BUCKETS_KEY + 
+      "All bucket sizes to be multiples of 256")}
\end{lstlisting}}
  \vspace{3.5pt}
  \centerline{\small (b) Add a semantic check (\href{https://github.com/apache/hbase/commit/bc93b6610b349d38502290af27da0ae0b5fd4936}{\HBASE-16993})}
  \vspace{5.5pt}
\end{minipage}
\begin{minipage}{0.97\linewidth}
  \centerline{
\begin{lstlisting}[language=diff]
+  require(conf.getOption(authKey).isEmpty()
+	   || !restServerEnabled,
+     s"The RestSubmissionServer does not " +
+      "support authentication via ${authKey}." +
+      "Either turn off spark.master.rest.enabled " +
+      "or do not use authentication.")
\end{lstlisting}}
  \vspace{3.5pt}
  \centerline{\small (c) Add a check for parameters dependency (\href{https://github.com/apache/spark/commit/10248758438b9ff57f5669a324a716c8c6c8f17b}{\SPARK-25088})}
  \vspace{5.5pt}
\end{minipage}
\begin{minipage}{0.97\linewidth}
  \centerline{
\begin{lstlisting}[language=diff]
+  if (rdd.isBarrier() && 
+      Utils.isDynamicAllocationEnabled(sc.getConf)) {
+    throw new SparkException(
+      "Barrier execution mode does not support"
+      "dynamic resource allocation for now. You can"
+      "disable dynamic resource allocation: setting"
+      "spark.dynamicAllocation.enabled to false")}
\end{lstlisting}}
  \vspace{3.5pt}
  \centerline{\small (d) Add a check for execution context (\href{https://github.com/apache/spark/commit/92b48842b944a3e430472294cdc3c481bad6b804}{\SPARK-24954})}
\end{minipage}
\caption{Examples of configuration checking code.}
\label{fig:code:CheckingCode}
\end{figure}

%
%

\subsubsection{Adding new checking code}
\label{sec:add:checking:code}

\checked{79} new checks were added in {\checked{83.6\% (56/67)}} of checking-code related commits.
\checked{87.3\% (69/79)} of these new checks were for specific parameters, while
the others were applied to groups of configuration parameters (e.g., read-only parameters).
Surprisingly, for specific parameter checks (\checked{69 checks in 46 commits}), 58.0\% (40/69) were\Space{ simple and} basic checks:
\CodeIn{NOT-NULL} \checked{(20/69)}, value range \checked{(15/69)} and deprecation checks \checked{(5/69)}.
An example\Space{ of such checks} is\Space{ shown} in Fig.~\ref{fig:code:CheckingCode}(a).
Majority of new checking code were added reactively,
corroborating that simple checks can
  prevent many severe failures~\cite{xu:16,yuan:14}. More of such checks could be
automatically added and invoked at system startup.
The other \checked{29} checks were more complex: \checked{9} value semantic checks
  (e.g., file/URI properties and data alignment, Fig.~\ref{fig:code:CheckingCode}(b)),
  2)~\checked{13} checks for parameter dependencies (e.g., Fig.~\ref{fig:code:CheckingCode}(c)),
and 3)~\checked{7} checks for execution context (e.g., Fig.~\ref{fig:code:CheckingCode}(d)).

\subsubsection{Improving existing checking code}

\checked{11} commits improved existing checking code:
\checked{eight}\Space{ of them} made checks more strict, e.g.,
a \CodeIn{NOT-NULL} check was improved to ``{\it only allow\Space{s} table replication
  for sync replication}''~(\href{https://issues.apache.org/jira/browse/HBASE-19935}{\HBASE-19935}), and
%
%
%
\checked{three} moved checking code to be invoked earlier instead
  of during subsequent execution, e.g., ``{\it \Space{when using dynamic
    executor allocation, if we set spark.executor.cores smaller than
  spark.task.cpus, an exception will be thrown.  But, if dynamic executor
  allocation not enabled, \spark{} will hang ...}\Space{ when submit new job as
  TaskSchedulerImpl will not schedule a task in an executor which
  available cores is small than spark.task.cpus. So, }when starting task
  scheduler, spark.task.cpus should be checked}''~(\href{https://issues.apache.org/jira/browse/SPARK-27192}{\SPARK-27192}).
\Space{We believe that checks should be performed earlier rather than later.}

\begin{shadedbox}
{\header{Discussion:}} Checks for parameter values
are often added as afterthoughts. Proactively generating checking code
can help prevent failures due to misconfigurations.
\end{shadedbox}

Two possible directions are automatically
learning checking code (we find that newly-added checking code is
often simple) and automatically applying checking code for one
parameter to other parameters both in the same software (which
developers are already doing manually) and across software
projects.
A direction is to co-learning checking code and usage code.
Techniques for extracting complex constraints
  and specifications can reduce
manual effort for reasoning about and implementing checking code.
A few recent works show promise for inferring parameter constraints through analysis
  of source code and documentation~\cite{xu:13,chen:2020,xiang:2020,Li:2020}.
  Techniques for extracting feature constraints could be extended and applied
  to runtime configurations~\cite{NadiTSE2015,nadi:14,kang:90,SheICSE11}.

\Comment{\haochen{(\it HHC: I inspected if there are data that can support this impl., but I fond this is a bit ``semantic'', depending on context. And some checks just can not move to startup stage.)}}


\begin{table*}[]
\begin{center}
\footnotesize
\caption{\small Four levels of message feedback quality in commits that changed exception or logging messages.}
\begin{tabular}{lll}
\toprule
\header{Level}    &   \header{Description}  &  \header{Example}        \\
\midrule
\multirow{2}{*}{L4}     &  Contain parameter names and &  ``Barrier execution mode does not support dynamic resource allocation... You can disable dynamic  \\
				    					  &  provide guidance for fixing   &  resource allocation by setting\Space{ \spark{} conf}...\ttvar{spark.dynamicAllocation.enabled} to false.''~(\href{https://github.com/apache/spark/commit/92b4884}{\SPARK-24954})  \\
\midrule
L3    & Contain parameter names &  ``Failed to create SSL context using \ttvar{server_encryption_options}.''~(\href{https://github.com/apache/cassandra/commit/16ef9ac}{\CASSANDRA-14991})     \\
\midrule
L2    & Do not contain parameter names    &   ``This is commonly a result of insufficient YARN configuration.''~(\href{https://github.com/apache/hbase/commit/2773510f}{\HBASE-18679})   \\
\midrule
L1    & No mention of configuration   &  ``Could not modify concurrent moves thread count.''~(\href{https://issues.apache.org/jira/browse/HDFS-14258}{\HDFS-14258})  \\
\bottomrule
\end{tabular}
\label{tab:msg_quality}
\end{center}
\end{table*}

\Comment{@Owolabi, I feel that misconfiguration correction is a direction to go.}


\subsection{Evolution of Error-Handling Code}
\label{sec:feedback:evolution}

We discuss changes to misconfiguration-related exception-handling code
and to messages that provide user feedback.

\subsubsection{Changes to configuration error handling}

\checked{19} commits dealt with error handling: \checked{10}
added new handling code to \CodeIn{try-catch} blocks
  or \CodeIn{throw} new exceptions; 9 commits changed handling code.
Among the 9 commits,
(1)~\checked{four} changed misconfiguration-correction code: three of these
added logic to handle a misconfiguration, e.g.
``{\it if secret file specified in httpfs.authentication.signature.secret.file does not exist,
random secret is generated}''~(\href{https://issues.apache.org/jira/browse/HDFS-13654}{\HDFS-13654}) and one\Space{, another commit do the opposite thing that developers} changed buggy
misconfiguration-correction code to simply log errors~(\href{https://issues.apache.org/jira/browse/HDFS-14193}{\HDFS-14193})
(showing that auto-correcting misconfigurations is not always easy),
(2)~\checked{three} changed the exception type as it was
``{\it dangerous to throw statements whose exception class does not
accurately describe why they are thrown...since it makes correctly handling
  them challenging}''~(\href{https://issues.apache.org/jira/browse/HDFS-14486}{\HDFS-14486}), and
(3)~\checked{two} replaced exception throwing with logging the errors and resuming the execution.

\Space{Besides the commits that changed error handling,}
  We also studied the newly added handling code in the \checked{79} commits
  that added new checking code in Section~\ref{sec:add:checking:code}\Comment{---what is the handling
  code when the newly-added checks fail?}.
In \checked{73.4\% (58/79)} of the cases,
the handling code threw runtime exceptions\Space{ with message} or logged error messages.
The expectation is that users should
  handle the errors.
In the remaining \checked{26.6\% (21/79)} cases, developers attempted to correct
  the misconfigurations, e.g.,
 ``{\it it's developers' responsibility
    to make sure the configuration don't break code.}'' (\href{https://github.com/apache/spark/pull/21601}{\SPARK-24610}).
Developers corrected misconfigurations by changing to the
  closest value in the valid range (\checked{11/21}),
  reverting to the default value (\checked{3/21}), and
  using the value of another parameter with similar semantics (\checked{7/21}).
%

\begin{shadedbox}
{\header{Discussion:}} \Space{We found that }Developers want to make code
more robust in the presence of misconfigurations, but their manual
efforts are often ad hoc. There is need for new techniques for
generating misconfiguration correction
code and improving existing handling code.
\end{shadedbox}
Techniques for fixing compile-time configuration errors,
  such as range fixes~\cite{xiong:12,xiong:15}, may be applicable
  for generating correction strategies for some types of runtime parameters.
A key challenge is to attribute runtime errors (e.g., exceptions) to
  misconfigurations and to rerun the related execution with the corrected configurations.


\subsubsection{Changes to feedback messages}
Feedback (error log or exception) messages
  are important for users to
  diagnose and repair misconfigurations\Space{ or to handle problems by tuning parameter values}.
We investigated commits that modified feedback messages and categorize the level of feedback
  that they provided in Table~\ref{tab:msg_quality}, where
  L4 messages provide the highest-quality feedback and L1 messages provide the lowest-quality feedback.
Among \checked{33} commits that modified messages, \checked{18}
  enhanced feedback quality by adding configuration-specific information.
After enhancement, \checked{8} messages became \CodeIn{L3}, and \checked{7} became L4.
%
Changes in the other 15 commits improved
1)~correctness (\checked{9/15})---half changed imprecise parameter
  boundary values, e.g., from ``no less'' to ``greater'' (\href{https://issues.apache.org/jira/browse/SPARK-26564}{\SPARK-26564}),
2)~readability (\checked{3/15}), such as fixing typographic errors\Space{ and rewriting sentences},
3)~the log level (\checked{2/15})\Space{ of the commits, 4/16 of the messages)}, and
4)~security (\checked{1/15}), i.e., removing potentially sensitive value.

\Comment{
How many of the 32 commits are about exception messages and how
  many are about logging during error correction? Also, why do we have 32 commits in this section but 34 in Table~\ref{tab:behavior}?
\needCheck{25} of the \needCheck{30} commits\Space{, 26/31 of the messages} involved exception messages\Space{ when problem happen. (e.g.
``{\it spark.network.timeout must be greater than the value of spark.executor.heartbeatInterval}'' ~\href{https://issues.apache.org/jira/browse/SPARK-26564}{\myminus \SPARK-26564})};\Space{ the other}
\needCheck{5}\Space{ of the 30 commits}\Space{, 5/31 of the messages are for} involved error-correction log messages.\Space{, e.g.(``{\it No HREGION\_COLUMNFAMILY\_FLUSH\_SIZE\_LOWER\_BOUND
set in table descriptor, using region.getMemStoreFlushSize of families instead.}'' ~\href{https://issues.apache.org/jira/browse/HBASE-19647}{\myminus \HBASE-19647})}
}


\begin{shadedbox}
{\bf \header{Discussion:}} Future work could study the
feedback level in \emph{all} messages related to misconfiguration handling code.
If most messages are not L4, then future work should
automatically detect deficient messages and
automatically enhance them to L4.
\end{shadedbox}


Moreover, configuration-related logging is not as mature as
logging for\Space{ regular} debugging~\cite{yuan:12, Yuan_icse12, asplos-logenhancer,
  yuan:10, Ding:15, Fu:15}. Improving configuration-related logging requires
logging related parameters, erroneous
values, and, where feasible, possible fixes.\Space{ Such future work is
needed:} Poor-quality feedback from tools
hinders developers~\cite{JohnsonEtAlICSE2013} and
techniques exist for dealing with
message errors in other domains~\cite{zhendong-su-finding-and-analyzing-compiler-warning-defects, BarikEtAlCompilerErrorsICSE14}.

\subsection{Evolution of Parameter Value Usage}

\Space{The way parameters and their values are used evolve over
time. }Software developers change how existing
parameters are used (``Use$_\text{Change}$'' in Table~\ref{tab:behavior})
and reuse existing parameters for different purposes
(``Use$_\text{Add}$'' in Table~\ref{tab:behavior}).

\Comment{This section also falls in short -- there are two aspects to look
  into: (1) do developers change existing uses (i.e., a parameter used
  in one way but changed into another)?
  (2) do developers reuse an existing parameter (i.e., add a new use)
  which is currently covered.}

\subsubsection{Changing how existing parameters are used}

\Comment{Actually ``use'' doesn't have a clear scope here.
So I will write what I have. Please help me to justify my mistakes.}
\Comment{I leave it to @Owolabi}

\checked{45 commits} changed parameter usage for
the following reasons:


\vspace{5pt}
\mypara{Fine-grained control.} In \checked{12/45} commits,\Space{14/41 params of the cases. The} developers previously used one parameter for multiple purposes,
 due to poor design---``e.g., {\it CompactionChecker and PeriodicMemStoreFlusher
   execution period are bound together\Space{, both use \CodeIn{hbase.server.thread.wakefrequency}}}'' (\href{https://issues.apache.org/jira/browse/HBASE-22596}{\HBASE-22596})---or
 for reuse---e.g., ``{\it \Space{spark.sql.execution.}arrow.enabled was added...\Space{ when we added} with PySpark\Space{ arrow optimization}...
Later, SparkR...\Space{ arrow optimization} was added...\Space{ and it's} controlled by the same parameter. Suppose users want to share some JVM between PySpark and SparkR...
They\Space{ are currently forced to} use the optimization for all or none.}'' (\href{https://issues.apache.org/jira/browse/SPARK-27834}{\SPARK-27834}).
Developers resolved both categories by creating separate parameters\Space{ (sometime will leave the original one) and} for fine-grained control.


\vspace{5pt}
\mypara{Domain/scope.} \checked{8/45} commits changed the
usage domain or scope of a parameter\Space{ (\needCheck{9/41 params})}.
For example, \hdfs{} developers changed a\Space{ the \ttvar{dfs.namenode.decommission.interval}} parameter, which was previously only used in the decommission phase to also be used in the maintenance phase, so ``{\it lots of code can be shared\Space{ between the\Space{ existing}
decommission functionality and to-be-added maintenance state support for datanodes}}'' (\href{https://issues.apache.org/jira/browse/HDFS-9388}{\HDFS-9388}).
\Comment{So . This param is used in DecommissionManager and
now this becomes DatanodeAdminManager. Both decommission and maintenance
will use this manager.}

\vspace{5pt}
\mypara{Parameter overriding}\checked{9/45} commits changed parameter override priority\Space{(\needCheck{7/41 params})}, e.g., ``{\it We need to support both table-level parameters. Users might also use session-level parameter\Space{ \CodeIn{\seqsplit{spark.sql.parquet.compression.codec}}}...\Space{
The priority rule will be like If other compression codec configuration was found
through hive or parquet,} the precedence would be...\Space{ \CodeIn{compression}, \CodeIn{\seqsplit{parquet.compression}},
\CodeIn{\seqsplit{spark.sql.parquet.compression.codec}}}}'' (\href{https://issues.apache.org/jira/browse/SPARK-21786}{\SPARK-21786}).

\vspace{5pt}
\mypara{Semantics}\checked{6/45} commits
changed what a parameter is used for\Space{(\needCheck{5/41 params})}, e.g., in
\href{https://issues.apache.org/jira/browse/SPARK-21871}{\SPARK-21871}, developers started using
\CodeIn{\seqsplit{spark.sql.codegen.hugeMethodLimit}} as the maximum\Space{
bytecode} compiled function size instead of
\CodeIn{spark.sql.codegen.maxLinesPerFunction}.

\vspace{5pt}
\mypara{Parameter replacement}\checked{6/45} commits
  swapped one parameter for another
  because the previous one was outdated or
  wrong \Space{(\needCheck{4/41 params})},
  e.g., in \href{https://issues.apache.org/jira/browse/SPARK-24367}{\SPARK-24367},
  a use of \CodeIn{\seqsplit{parquet.enable.summary-metadata}} was replaced with
  a use of \CodeIn{\seqsplit{parquet.summary.metadata.level}} because the former was deprecated.

\vspace{5pt}
\mypara{Buggy parameter values}\checked{4/45}
commits changed parameter values that were buggy\Space{ (\needCheck{2/41 params})},
e.g., the value of a parameter\Space{ \CodeIn{\seqsplit{spark.ui.filters}}}
changed because, ``{\it user specified filters are not applied in YARN
mode\Space{, as it is overridden by the yarn AmIp filter. So}...we need...\Space{ to append} user provided filters\Space{ with yarn filter}}'' (\href{https://issues.apache.org/jira/browse/SPARK-26255}{\SPARK-26255}).


\Comment{
using \ttvar{spark.sql.codegen.maxLinesPerFunction} to restrict the maximum lines of a single Java
function generated by whole-stage codegen. Developers now use \ttvar{spark.sql.codegen.hugeMethodLimit}
to restrict The maximum bytecode size of a single compiled Java function generated by whole-stage codegen.
However. Both of those tow cases change the parameter's name after changing its functional usage.
}

\subsubsection{Reusing existing parameters}
\label{sec:reuse:parameters}

To avoid growing the configuration space unnecessarily, developers
sometimes reuse existing parameters that are similar to their new use
case, instead of introducing a new parameter. \checked{110}
commits reused \checked{151} existing parameters for
different purposes. However, parameter reuse comes at a cost.
We find two main problems. First, reusing a parameter and code
that it controls can result in subtle inconsistencies that can lead to bugs
or user confusion.
\checked{19.2\% (29/151)} parameter reuses had such inconsistencies.
Second,\Space{ even when there are no
inconsistencies,} developers often\Space{ resort to} clone existing code to enable
reuse.\Space{ In the rest of this section, } We focus on
inconsistencies. Problems
of code cloning are the subjects of other
research.

\Space{To find inconsistencies in parameter reuse we manually compared the old
and new uses in the commit where we saw the reuse. First we
  check the instant version \yuanliang{this is a bad word, I mean the
    version of this commit} of the commit that use existing
  parameter. And try to find whether there are inconsistency between
  the new use and old use. Finally, we find \needCheck{22} inconsistent code
  behavior among \needCheck{125} parameter new usage. (17.6\%).}

We manually checked for inconsistencies by comparing the newly-added code in a target commit
  with code that used the parameter in \blue{existing code base}.
We found \checked{29} inconsistencies in \hdfs{} \checked{(9/29)}, \hbase{}
\checked{(9/29)} and \spark{} \checked{(11/29)}.\Space{ We did not find inconsistent
parameter reuse in Cassandra.}
Inconsistencies manifest in different ways.
We categorized them based on the sources of inconsistencies during reuse:
  1) feedback message \checked{(9/29)}, e.g.,~\href{https://issues.apache.org/jira/browse/SPARK-18061}{\SPARK-18061};
  2) checking code \checked{(4/29)} e.g.,~\href{https://issues.apache.org/jira/browse/HBASE-20590}{\HBASE-20590};
  3) new uses of deprecated parameters \checked{(6/29)}, e.g.,~\href{https://issues.apache.org/jira/browse/HDFS-12895}{\HDFS-12895};
  4) default values \checked{(3/29)}, e.g.,~\href{https://issues.apache.org/jira/browse/HBASE-21809}{\HBASE-21809};
  and 5) use statements \checked{(7/29)}, e.g.,~\href{https://issues.apache.org/jira/browse/HBASE-20586}{\HBASE-20586}.
Fig.~\ref{fig:code:InconsistentUse} shows examples of inconsistencies in reuse of checking code
and use statements, where added lines start with $+$.
In Fig.~\ref{fig:code:InconsistentUse}(a), the new parameter usage did not check for parameter value emptiness as the old usage did.
In Fig.~\ref{fig:code:InconsistentUse}(b), the new usage of \CodeIn{hbase.security.authentication} checked case-insensitive equality; the old usage was case-sensitive.


\begin{shadedbox}
{\header{Discussion:}} Inconsistencies in
parameter usage can confuse users (the same values are
used in different ways) or lead to bugs. Ideas for detecting bugs as deviations from
similar program behavior~\cite{engler:01,tan:08} could
be starting points for addressing this problem.
\end{shadedbox}

\begin{figure}
\begin{minipage}{0.97\linewidth}
\centering
  \centerline{
\begin{lstlisting}[language=diff]
+  String principal = conf.get(
+	  Constants.REST_KERBEROS_PRINCIPAL);
+   if (principal != null) {...}

Preconditions.checkArgument(principalConfig != null
	&& !principalConfig.isEmpty(),
	REST_KERBEROS_PRINCIPAL + 
	" should be set if security is enabled");
\end{lstlisting}}
  \vspace{3.5pt}
  \centerline{\small (a) Inconsistent checking (\href{https://github.com/apache/hbase/commit/7da0015a3b58a28ccbae0b03ba7de9ce62b751e1}{\HBASE-20590})}
  \vspace{3.5pt}
\end{minipage}
\begin{minipage}{0.97\linewidth}
\centering
  \centerline{
\begin{lstlisting}[language=diff]
+  if(peerConf.get("hbase.security.authentication")
+      .equals("kerberos")) {...}

isSecurityEnabled = "kerberos".equalsIgnoreCase(
    conf.get("hbase.security.authentication"));
if (isSecurityEnabled) {...}

\end{lstlisting}}
  \vspace{3.5pt}
  \centerline{\small (b) Inconsistent parameter usage (\href{https://github.com/apache/hbase/commit/cd61bcc01eb29eb3509f72cf72326605afefabc8}{\HBASE-20586})}
  \vspace{3.5pt}
\end{minipage}
\caption{Examples of configuration inconsistent reuse.}
\vspace{-10pt}
\label{fig:code:InconsistentUse}
\end{figure}

%

\subsection{Summary}

\Space{Based on our findings on configuration usage evolution,}
We advocate that improving software qualities---resilience, diagnosability,
  and consistency---should be
  first-class principles in software configuration engineering.
We find that even in mature,
  production-quality cloud systems,
  checking, error handling, feedback, and parameter usage are often not designed
  or implemented in a principled manner.
More research effort should be put on enhancing these essential qualities
  of configurable software to defend against misconfigurations,
  in addition to detection
 and diagnosis tools that are external to the cloud system~\cite{attariyan:10,
   attariyan:12,attariyan:08,zhang:13,zhang:14,
   wang:04,wang:03,santo:16,santo:17,dong:15,
   tang:15,huang:15,Baset:2017,sun:osdi:20}.

\section{Configuration$\:$Document$\:$Evolution}
\label{sec:document}

We very briefly discuss configuration document evolution:
\checked{114} commits made \checked{149} changes to user manuals or code comments.
100 of these commits changed user
  manuals and the rest\Space{
  \checked{12.3\% (14/114)}} changed code comments.
We discuss why configuration documents were changed and the changed content.



\vspace{5pt}
\mypara{\bf Reasons for changing configuration documents.}
The \checked{149} changes\Space{ (in \checked{114} commits)} to configuration documents resolved five types of
  problems\Space{ (and their proportions): previous information was}:
 1)~\checked{63} were inadequate\Space{ \checked{(63/149)}} for users to understand parameters or
    to set values correctly,\Space{. A developer making one such change in \spark{} said} e.g.,
    ``{\it\Space{ Couple of }users wondered why spark.sql.shuffle.partitions...unchanged when they changed the config\Space{ value after running the query. It's}...worth to explain it in guide doc}'' (\href{https://issues.apache.org/jira/browse/SPARK-25245}{\SPARK-25245});
 2)~\checked{29} were outdated\Space{ \checked{(29/149)}}} after
    \scdi{} changed (Section~\ref{sec:interface:evolution}
      and Section~\ref{sec:behavior:evolution});
 3)~\checked{21} were incorrect\Space{ \checked{(21/149)}}\Space{Developers found that the previous information was not correct. An example from \hdfs{}:}, e.g., ``{\it LazyPersistFileScrubber will be disabled if\Space{ scrubber
 interval is}... configured to zero. But the document was incorrect\Space{---setting it to a negative value to disable this behavior}}'' (\href{https://issues.apache.org/jira/browse/HDFS-12987}{\HDFS-12987});
 4)~\checked{17} had readability issues\Space{ \checked{(17/149)}}, e.g.,
 ``{\it Client rpc timeouts are not easy to understand from\Space{ the} documentation}'' (\href{https://issues.apache.org/jira/browse/HBASE-21727}{\HBASE-21727}); and
 5)~\checked{19} improved content\Space{ \checked{(19/149)}}\Space{. The developer try to extend the document by adding configuration contents.},
 e.g., ``{\it Add thrift scheduling\Space{ pool}... config to scheduling docs}'' (\href{https://issues.apache.org/jira/browse/SPARK-20220}{\SPARK-20220}).

\begin{shadedbox}
{\bf \header{Discussion:}}
Document-as-code techniques can be applied to eliminate inconsistencies between
  configuration documents and configuration design/implementation.
\end{shadedbox}

\vspace{5pt}
\mypara{Content added to enhance documents.}
Inadequate information was the most common problem resolved by
configuration document changes.
We put the \checked{63} changes that enhanced inadequate documents in
  six categories \Comment{(and their proportion)}based on the content added:\Comment{
  were added by developers, as an indication of the types of information
  that should be provided in the first place.}
1)~\checked{16} changed constraints on parameter values\Space{ \checked{(16/63)}},
  e.g., ``{\it This should be positive and less than 2048}'' (\href{https://github.com/apache/cassandra/commit/a586f6c88dab173663b765261d084ed8410efe81}{\CASSANDRA-13622});
2)~\checked{10} explained dependence on other parameters\Space{ \checked{(10/63)}},
  e.g., ``{\it This property works with dfs.namenode.invalidate.work.pct.per.iteration}''
    (\href{https://github.com/apache/hadoop/commit/b0560e0624756e2b3ce7b6bc741eee3c18d2a873}{\HDFS-12079});
3)~\checked{6} changed parameter value types and units\Space{ \checked{(6/63)}};
4)~\checked{6} changed parameter scope\Space{ \checked{(6/63)}}, e.g.,
   ``{\it Timeout\Space{ for scan operations}... is controlled differently. Use hbase.client.scanner.timeout.period
   property to set this timeout}'' (\href{https://github.com/apache/hbase/commit/51c58e083ca89a33de79c8531a16f7072c488d6d}{\HBASE-21727}),
5)~\checked{22} provided use cases and guidance\Space{ \checked{(22/63)}}, e.g.,
\Space{  \texttt{\small dfs.image.compress} in}
  ``{\it enabling this will be very helpful if dfs image is large}'' (\href{https://issues.apache.org/jira/browse/HDFS-13884}{\HDFS-13884}); and
6)~\checked{3} warned about deprecation,
  e.g., ``{\it this config will be removed in Spark 3.0}''~\href{https://issues.apache.org/jira/browse/SPARK-25384}{(\SPARK-25384)}.
\Space{We recommend these categories of information to be systematically documented and maintained.}

\Comment{
\begin{table}
\scriptsize
\caption{Statistics of configuration document evolution.}
\begin{tabular}{lccccc}
\toprule
 & \header{HDFS} & \header{HBase} & \header{Spark} & \header{Cassandra} & \header{total} \\
\midrule
   {\bf User manual}             	& \checked{27} & \checked{18} & \checked{52} & \checked{3} & \checked{100} \\
   {\bf Code comments}             & \checked{0} & \checked{3} & \checked{9} & \checked{2} &  \checked{14} \\
\bottomrule
\end{tabular}
\label{tab:Documentation}
\end{table}
}

\begin{shadedbox}
{\bf \header{Discussion:}}
Ethnographic studies could help understand the gaps between documented
configuration information and configuration obstacles faced by users.
\end{shadedbox}

\vspace{3pt}
\mypara{Summary}
Correctness and effectiveness of technical documentation is a long-lasting
  problem in software engineering.
Configuration documentation is no exception.
Specialized techniques for maintaining and improving configuration documentation are needed.
For example, checking for inconsistencies between documents
  and source code~\cite{tan:07,tcomment,yzhou:18,hzhong:13}
  could help detect defects in configuration-related code
  or documents.
Also, techniques for auto-generating documents, especially
  using structured data,
  can be applied to generating per-parameter
  comments and manual entries~\cite{autocomment,cpc,Giriprasad:10}.

\section{Threats to Validity}

We studied cloud systems.
Some of our findings may not generalize to other kinds of software\Space{ (e.g. mobile applications, embedded software)}.
We chose these projects because they are widely used, highly configurable
with lots of parameters, mature, and well maintained.
They also have issue-tracking systems that help us understand the context of \confcommits.

Though we selected candidate commits from version control history, we may have missed some \confcommits{} due to two limitations.
First, our regular expressions assume standard coding conventions and will not match if developers do not follow these conventions.
Second, our simple, text-based tainting may miss some changes to the data flow of variables that store parameter values.
However, as we mentioned in Section~\ref{sec:meth}, precise pairwise tainting does not scale to all the commits in the range that we studied---we traded off precision for scalability.
All commits selected were manually inspected and categorized through a rigorous
  quality-assurance process (Section~\ref{sec:inspection_cat}).
\Space{ the simple approach recognizes configuration-related diff,
the} False positives came mainly from commits that touched lines containing configuration-related variables but
did not change the configuration.


\Comment{CASSANDRA-13578, simplify mx4j configuration,2017-06-07}


\section{Related Work}
\label{sec:relatedwork}

\Space{We presented one of the first studies that systematically analyze
  different dimensions in the evolution
  of software configuration design and implementation.}
A prior study~\cite{sai:14} found that software evolution necessitates
  resetting parameter values and\Space{ the
  usage of configuration parameters; thus, when
  upgrading to a new software version,
  users may need to re-configure the software with different parameter values.
They} built \CodeIn{ConfigSuggester} to identify parameters whose values need to be
  changed after a software updates. We study how the configuration interface and parameter
  usage change across (a portion of) version control history to draw insights for
  better \scdi{}.
Sayagh et al.~\cite{sayagh18} studied
  software configuration engineering in practice using interviews,
  user surveys, and a literature review.
Our work is complementary: we perform a \emph{code}-level study of configuration\Space{
  engineering with respect to software} evolution, which\Space{.
Our analysis of code-level configuration evolution
  from developers' experiences and practices} yields new insights. 


There have been many studies
  on misconfigurations in a wide variety of software
  systems~\cite{amvro:16,tang:15,barroso2018,yin:11,kendrick:12,
    rabkin:13,gunawi:16,maurer:15,
    oppenheimer:03,nagaraja:04}.
Our work does not focus on detecting misconfigurations
    or diagnosing failures caused by misconfigurations.
    We focus on current configuration engineering practices,
    with the goal to understand how to improve configuration design and implementation.
\Space{
    on reducing misconfigurations (as well as other configuration
    problems) in the first place, with better
    design and implementation.}

Recently, a few studies investigated automated techniques or
      engineering practices to enhance configuration
      checking code~\cite{xu:16},
      diagnosability~\cite{zhang:15},
      interface~\cite{xu:15}, security~\cite{Meng:18,xiang:19,xu:chi:17},
      configuration data analysis~\cite{xu:18},
      configuration libraries~\cite{SayaghSCAM,raab:17},
      and correlations or coupling in
      configuration and code~\cite{Horton:19,Wen:20,mehta2020rex}.
Our work corroborates and complements the aforementioned work from the perspective
  of software evolution. Specifically, our work studies the practices of
  {\it software developers} and reveals
  how software configuration design and implementation are revised and evolved.


Despite the differences (Appendix A), \Space{software} runtime configurations share commonalities
  with compile-time configurations or SPL configurations, such as \CodeIn{\#ifdef}-based feature
  flags~\cite{meinicke:20}. It is possible that techniques and methodologies designed for compile-time
  configurations, especially feature and variability modeling~\cite{Lotufo:10,Passos:18,Berger2013TSE,Damir2019FSE,Berger:10,Liebig:10},
  could be adapted for use with runtime configurations.
Such adaptation needs to address unique challenges of runtime configuration parameters,
  such as dependencies on deployment environments, as well as their complex data types
  and misconfiguration patterns.


Configuration design and implementation have significant implications
  on software testing and debugging~\cite{yilmaz:06,jin2:14,fouche:09,qu:08,reisner:10,song:12}.
For example, introducing new parameters
  enlarges the configuration space and thus makes it more costly to comprehensively
  test software.
In this paper, we focus on understanding how to improve \scdi{} so that fewer misconfigurations occur,
  and not on software bugs that can occur under different parameter value combinations.\Space{
Understanding the impact of configuration design/implementation on software code quality is our future work.}

\Comment{
We limit our study scope to deployment-time parameters that can be
  changed via configuration files or CLIs.
We do not study build-time configurations studied in the variability
  and software product line research.
Our work is also complementary to configuration-aware combinatorial testing.
}


\section{Conclusions}
\label{sec:conclusions}

We presented present an evolutionary study of
  configuration design and implementation in cloud systems. To the best of our knowledge, ours is the first evolutionary study on code-level
  runtime \scdi{} in these systems.
We analyze rationales and practices for revising \scdi{}
decisions, especially in response to consequences of
misconfigurations. Our study yields several new insights into
the configuration engineering process,
and research opportunities for
reducing misconfigurations.
Our hope is to inspire researchers and developers to treat
  configuration engineering
  as a first-class software engineering endeavor.


\section*{Appendix A: Runtime versus SPL Configuration}
\label{sec:appendix}

A very frequent request is to compare runtime configuration (the type of
  configuration studied in this paper) with software product lines (SPL) configuration
  (often referred to as ``feature flags'' or ``feature toggles'')
  and to position the work in the area of SPL
  and variability modeling. We explicitly discuss a few fundamental differences:

First, runtime configurations are changed by software users (operators/sysadmins
  in our context); SPL configurations are managed by developers.
  Since users are unfamiliar with code, the configuration specifications
  become the interfaces (Section~\ref{sec:interface:evolution}).
Moreover, as users are prone to misconfigurations, checking and providing feedback are critical (Section~\ref{sec:behavior:evolution}).

Second, runtime configurations are implemented differently than SPL configurations.
  Runtime configurations are mostly in the form of configuration parameters
  that load values from files or command lines; SPL
  configurations are typically in the form of preprocessors
  that determine modules to be included in the released binary.

Third, runtime configurations of cloud software are changed frequently
  (hundreds to thousands of times a day~\cite{tang:15,mehta2020rex,maurer:15});
SPL configurations are typically changed with product release cycles.
This higher velocity of runtime configuration changes increases
  misconfiguration occurrences and makes checking, error handling, and logging critical.

Fourth, runtime configurations depend on the deployment environment,
  including machine resources (e.g., CPU, memory, and storage),
  operating systems (e.g., files, IP addresses, and ports),
  and workloads (data size and requests per seconds). In contrast, SPL configurations
  are often determined before software release or system deployment.

Lastly, runtime configurations have more complex data types (e.g., string and numeric)
  with different error patterns; SPL configurations are mostly boolean or enumerative types.

Certainly, ideas in SPL and variability modeling can be extended and applied
  to runtime configuration.
We have discussed them in context of our analysis throughout the paper.

\section*{Appendix B: Replication Package}

We release our research artifacts of this paper at:
\begin{center}
\fbox{\url{https://github.com/xlab-uiuc/open-cevo}}
\end{center}
The artifacts include:
\begin{itemize}
  \item the script code for collecting commits and JIRA issues;
  \item the annotated dataset of each categories (interface, usage, and document);
  \item documents for running the code and navigating the data.
\end{itemize}

We believe our study can be reproduced by different teams based on the taxonomy
  described in
  Figure~\ref{sec:overview} and Table~\ref{tab:taxonomy}, together with the
  methodology described in Section~\ref{sec:meth}.

\vspace{5pt}
\noindent
{\bf Author experience.} All the authors have been working on software configuration
  research for several years, ranging from two to nine years.
The authors have a good understanding of the four systems under study (HDFS, HBase,
  Spark, and Cassandra)---they have used those
  studied systems as subjects of evaluation in their prior research.

We expect that a similar level of experiences and expertises are needed for a team to
  reproduce the analysis (mainly categorization), including the understanding
  of the designs of the systems under study and the implementations
  in Java/Scala programming languages to understand
  the evolution. We believe a fair understanding of software configuration
  design and implementation is required, too.

\vspace{5pt}
\noindent
{\bf Heuristics for commit identification and analysis.} The heuristics are
  described in Section~\ref{sec:diff:analysis}. The script code that implements the heuristics
  can be found at:
  \begin{center}
  \fbox{\url{https://github.com/xlab-uiuc/open-cevo/tree/main/code}}
  \end{center}
We documented how to run the code and the basic regular expressions used at the \texttt{README.md} file.

\vspace{5pt}
\noindent
{\bf Identification/categorization of the rationales for parameterization.}
We identify and categorize the rationales for parameterization based on the descriptions
  (from the reporter) and the discussions (among the developers) documented in the JIRA issues.
We developed our categories bottom-up instead of imposing a taxonomy {\it ex ante}.
Specifically, we list all the rationales/reasons and then summarizing them into categories.
We do not claim completeness of the categories.

We present an example from Spark, in which
the commit (\href{https://github.com/apache/spark/pull/21705/commits/8ee8e00f1}{link}) changed a constant value \texttt{100}
to a configuration parameter, \texttt{spark.sql.codegen.cache.maxEntries};
the commit is associated with the JIRA issue,~\href{https://issues.apache.org/jira/browse/SPARK-24727}{SPARK-24727}.
Quoting the JIRA issue description,
\begin{quote}
``{\it The cache 100 in CodeGenerator is too small for realtime streaming calculation, although is ok for offline calculation.
Because realtime streaming calculation is mostly more complex in one driver, and performance sensitive.
I suggest spark support config for user with default 100, such as spark.codegen.cache=1000.}''
\end{quote}
Based on the description, we conclude that the rationale behind the parameterization is that the constant value
  cannot meet the performance requirement of real-time application.
So, it is categorized as ``performance'' in Table~\ref{tab:purpose_constant2config}.

\vspace{5pt}
\noindent
{\bf Identification on how developers find constants to parameterize.}
It is straightforward to identify the constants that were parameterized in a selected commit.
The following commit (\href{https://github.com/apache/spark/commit/565e7a8d}{link}) illustrates this point; the commit comes
  from ~\href{https://issues.apache.org/jira/browse/SPARK-20950}{SPARK-20950}.
We can see that the constant (\texttt{1024*1024}) was parameterized---the value
  was replaced by a program variable
  \texttt{diskWriteBufferSize} read from a configuration parameter.
\begin{lstlisting}[language=diff]
- int DISK_WRITE_BUFFER_SIZE = 1024 * 1024
+ int diskWriteBufferSize =
+   conf.get(package$.MODULE$.
+	    SHUFFLE_DISK_WRITE_BUFFER_SIZE)
- byte[] writeBuffer = 
-     new byte[DISK_WRITE_BUFFER_SIZE]
+ byte[] writeBuffer = 
+     new byte[diskWriteBufferSize]
\end{lstlisting}



For each commit that parameterized constants, we identify the intent of the parameterization,
  i.e., ``how developers find constants to parameterize.''
Similar to the identification/categorization of the rationales for parameterization,
 we identify and categorize the reasons for how developers find constants to parameterize
  based on the descriptions (from the reporter) and the discussions (among the developers)
  documented in the JIRA issues. The process is identical---we list all the rationales/reasons and then summarizing them into categories.
  We do not claim completeness of the categories.

We present an example from Spark, in which the commit (\href{https://github.com/apache/spark/commit/126310ca}{link})
  parameterized the thread number of broadcast-exchange thread pool from a constant \texttt{128}.
Based on the associated JIRA issue, \href{https://issues.apache.org/jira/browse/SPARK-26601}{SPARK-26601}:

\begin{quote}
``{\it Currently, thread number of
 broadcast-exchange thread pool is fixed and keepAliveSeconds is also fixed as 60s. But some times,
 if the thread object do not GC quickly it may caused server(driver) OOM. In such case, we need to make this thread pool configurable.}''
\end{quote}
Therefore, we conclude that the parameterization is to avoid OOMs, a type of ``failures'' (Section~\ref{sec:parameterization}).

\vspace{5pt}
\noindent
{\bf Identification/categorization of the reasons for changing default values.}
Similar to the identification/categorization of the rationales for parameterization,
  we identify and categorize the reasons for changing default values
  based on the descriptions (from the reporter) and the discussions (among the developers)
  documented in the JIRA issues. The process is identical---we list all the
  rationales/reasons and then summarizing them into categories.
  We do not claim completeness of the categories.

We present an example from HDFS. The commit (\href{https://github.com/apache/hadoop/commit/afb42aeab}{link})
  changed the default value of \texttt{dfs.namenode.edits} \texttt{.asynclogging} from \texttt{off} to \texttt{on}.
The commit is associated with the JIRA issue,~\href{https://issues.apache.org/jira/browse/HDFS-12603}{\HDFS-12603},
  which in turn was linked to~\href{https://issues.apache.org/jira/browse/HDFS-7964}{\HDFS-7964}.
Quoting the discussion:
\begin{quote}
``{\it It was off by default due to concerns about correctness.
We have been running it in production for quite a while with no issues so far.}''
\end{quote}
We conclude that the commit enabled a previously-disabled feature flag (Section~\ref{sec:default_value}).

\vspace{5pt}
\noindent
{\bf Definitions of four levels of messages and feedback quality.}
The definitions are developed based on how the messages are changed, i.e.,
  how developers improved the original messages.
Specifically, we summarized what additional information is added to the original
  message, which can be categorized into the four levels.
The following is a commit (\href{https://github.com/apache/hbase/commit/4b84ab32}{link})
  that changed the message which improved the message quality from L3 to L4
  defined in Table~\ref{tab:msg_quality}.

\begin{lstlisting}[language=diff]
-  resp.getWriter().write(
-	 "ASYNC_PROFILER_HOME is not set.");
+  resp.getWriter().write(
+	 "ASYNC_PROFILER_HOME is not set." +
+    "Please ensure prerequsites for the Profiler" + 
+    "Servlet have been installed and the" +
+    "environment is properly configured." + 
+    "For more information please see" +
+    "http://hbase.apache.org/book.html#profiler");
\end{lstlisting}

\vspace{5pt}
\noindent
{\bf Handling commits that change multiple categories.}
One commit could revise multiple categories in our taxonomy (Figure~\ref{sec:overview} and Table~\ref{tab:taxonomy}).
In total, there are 26 commits that revise multiple categories.
For those commits, we study them in each category independently.
For example, the following commit (\href{https://github.com/apache/spark/commit/10248758}{link})
changes the default value of \texttt{spark.master.rest.enabled};
meanwhile, it also adds the checking code for \texttt{SPARK\_AUTH\_SECRET\_CONF}.
\begin{lstlisting}[language=diff]
-  val restServerEnabled = conf.getBoolean(
-    "spark.master.rest.enabled", true)
+  val restServerEnabled = conf.getBoolean(
+    "spark.master.rest.enabled", false)
...

+  authKey = SecurityManager.SPARK_AUTH_SECRET_CONF
+    require(conf.getOption(authKey).isEmpty 
+	    || !restServerEnabled,
+       s"The RestSubmissionServer does" + 
+        "not support authentication.")
\end{lstlisting}
Therefore, we study this commit in both Mod$\rm_{\text{DefaultValue}}$
  (Section~\ref{sec:default_value}) and Check (Section~\ref{sec:usage:check}).

\section*{Acknowledgement}

We thank Xiangbing Huang, Xudong Sun, Sam Cheng, Jack Chen, and Darko Marinov
  for discussions.
The research was mainly conducted when Zhang was a visiting student at
  UIUC, supported by China Scholarship Council.
Zhang, He, Li, and Dong were supported in part of National Key R\&D Program of
  China No. 2017YFB1001802; NSFC No. 61872373 and 61872375.
Xu was supported in part of NSF 1816615.

\bibliographystyle{ieeetr}
\bibliography{ref,ref3,ref-ICSE20SEIP}

\end{document}